\documentclass[aps,prd,11pt,a4paper,nofootinbib,oneside,superscriptaddress]{revtex4-1}
\pdfoutput = 1

\usepackage{amsmath,amssymb,amsfonts,color}
\usepackage{tensor,slashed,paralist,cases,mathrsfs}
\usepackage{float,cancel,xcolor}
\usepackage{graphicx}
\usepackage{dcolumn}
\usepackage{bm}

\usepackage[colorlinks=true,urlcolor=blue,linkcolor=blue,citecolor=blue,linktocpage=true]{hyperref}

\begin{document}

\vspace*{1.5em}

\title{Fermionic Dark Matter through a Light Pseudoscalar Portal: Hints from the DAMA Results}

\author{Kwei-Chou Yang}
\email[Electronic address: ]{kcyang@cycu.edu.tw}

\affiliation{Department of Physics and Center for High Energy Physics, Chung Yuan Christian University, Taoyuan 320, Taiwan}


\begin{abstract}

We study the fermionic dark matter (DM) particle interacting with Standard Model quarks via a light pseudoscalar mediator. We consider separately the scenarios for which the DM-pseudoscalar coupling is $CP$ conserving or $CP$ violating.
We show that taking a contact interaction is not suitable, even when the mediator has a mass of the same order of magnitude as the typical momentum transfer  at the direct-detection experiments, such that the allowed DAMA region is excluded or considerably modified by the correct relic density requirement.
The DAMA result seems to indicate that the $CP$-violating interaction is dominant at direct searches. We find that, if the proton-to-neutron effective coupling ratio is  $-60\sim -40$, the exclusion limits set by  SuperCDMS, XENON100, and LUX are highly suppressed, and the DAMA signal can thus be easily reconciled with these null measurements. 
For this model, the allowed region determined by the DAMA signal and correct relic density can successfully satisfy the  conditions required by the thermal equilibrium, big bang nucleosynthesis, and DM self-interactions.  The results of future measurements on flavor physics will provide important constraints on the related models. Precise measurements performed by COUPP, PICASSO, SIMPLE and KIMS should be able to test this model in the near future.

\end{abstract}
\maketitle
\newpage

\section{Introduction}

The evidence for existence of dark matter (DM) in the Universe has been established  by various astronomical observations and astrophysical measurements. Our Milky Way galaxy is believed to be surrounded by a halo of the DM whose composition, however, remains unknown. It was suggested that the DM may be composed of the so-called weakly interacting massive particles (WIMPs) with mass of order $10- 10^3$ GeV, which can connect  with new-physics phenomenology at the electroweak scale.  The WIMPs can easily model the relic abundance, matching the observed (cold) dark matter density   $\Omega_{\rm{DM}}  h^2=0.1198 \pm 0.0026$ \cite{PDG,Ade:2013zuv}.

The direct-detection searches from DAMA \cite{ Bernabei:2010mq,Bernabei:2013xsa}, CoGeNT \cite{COGENT}, CRESST-II \cite{CRESSTII}, and CDMS-Si \cite{CDMS-Si} have observed an excess number of events over the background in their counting rates; DAMA in particular has claimed to observe events at a very high significance of 9.3$\sigma$. These results have been interpreted as evidence for DM with a mass ${\cal O}(10)$ GeV. However, these results are not supported by the null measurements  \cite{XENON100,SuperCDMS,COUPP,PICASSO,LUX,Felizardo:2011uw,Kim:2012rza}. DM interactions with the nucleus through the ordinary spin-independent and spin-dependent operators, which at the quantum level are independent of  momentum transfer and of relative velocity, have been well studied. Nevertheless, such theoretical predictions did not explain the experimental anomalies. 
The momentum-dependent DM interaction with ordinary matter mediated by a pseudoscalar coupling were then motivated \cite{Chang:2009yt, Freytsis:2010ne, Fan:2010gt, Tsai:2013bt, Fedderke:2014wda, Ghorbani:2014qpa, Berlin:2015wwa, Kim:2016csm,Dev:2015isx}, but the related response form factors were not well studied until 2012 \cite{Fitzpatrick:2012ix,Anand:2013yka}. The most general nonrelativistic (NR) effective theory for one-body dark matter-nucleon interactions was stressed recently in  Refs.~\cite{Fitzpatrick:2012ix,Anand:2013yka}, where the relevant nuclear response form factors for each of the NR operators were computed.

 From analyzing the Fermi Gamma-ray Space Telescope (Fermi-LAT) data, several studies have found a GeV gamma-ray excess arising from the region of the Galactic center (GC) \cite{Goodenough:2009gk, Hooper:2010mq, Hooper:2011ti, Abazajian:2012pn, Gordon:2013vta, Huang:2013pda, Daylan:2014rsa, Calore:2014nla}.
Boehm et al. \cite{Boehm:2014hva} have shown that the Dirac fermionic (Coy) DM $\chi$ with mass $m_\chi\sim 30$~GeV, interacting with Standard Model (SM) particles $f$ via a pseudoscalar mediator $A$
 with $CP$-conserving Yukawa-like couplings ($g_p^f \propto m_f/v$),
 \begin{eqnarray}\label{eq:ex-lagrangian-1}
{\cal L}_{\rm int}\supset  g_{p,\chi} A \overline{\chi}  
i  \gamma^5  \chi +  i A \sum_f g_p^f \bar f  \gamma^5 f \,,
\end{eqnarray} 
 can explain the GC gamma-ray excess, and result in a strong suppression in other experimental searches. 
 It was shown that the forthcoming run of the LHC can constrain regions of the parameter space where  $m_A>2 m_\chi$ \cite{Boehm:2014hva,Buchmueller:2015eea};
 see also extensive works in Refs.~\cite{Hektor:2014kga,Izaguirre:2014vva,Ipek:2014gua, Arina:2014yna,Abdullah:2014lla,Cheung:2014lqa,Huang:2014cla,Cahill-Rowley:2014ora,Dolan:2014ska,Kozaczuk:2015bea}.

    In this paper, taking a bottom-up approach, we will consider the fermionic DM particle interacting with the SM quarks via a light pseudoscalar mediator with a $CP$-conserving ${\bar\chi}$-$\chi$-$A$ coupling described by the Lagrangian in Eq.~(\ref{eq:ex-lagrangian-1}) and, separately, with a $CP$-violating ${\bar\chi}$-$\chi$-$A$ coupling ${\cal L}_{\rm int}\supset   g_{s,\chi}    A \overline{\chi}  \chi$.  Taking into account the correct response form factors, the latter has not been previously systematically studied. Although this $CP$-violating interaction, for which the DM annihilates through $p$-waves into SM quarks, is not suitable for explaining the GC gamma-ray excess, the unresolved millisecond pulsars may instead be responsible for the excess observation \cite{Abazajian:2014fta}.
We perform a detailed study of the couplings and the mediator's mass that may be allowed by current experimental constraints. 
 Naively, one would expect $g_{s,\chi} \ll  g_{p,\chi} $ due to the smallness of the $CP$-violating effect.  Interestingly, the $CP$-violating interaction could be even stronger than the $CP$-conserving one in the direct-detection experiments if  $g_{s,\chi} / g_{p,\chi}  >|\vec{q}|/(2m_\chi)$ \cite{Dienes:2013xya,Dolan:2014ska}, where $\vec{q}$ is the momentum transfer in the DM-nucleus scattering.  We will show that the $CP$-violating interaction can offer a good fit to the DAMA data  \cite{ Bernabei:2010mq,Bernabei:2013xsa}, where $|\vec{q}|/(2m_\chi)\sim 10^{-3}$ is satisfied.

We reexamine the DAMA modulation signal and focus on the phenomenology related to the light mediator  with mass $m_A <m_\chi$, especially in the sub-GeV region. This is motivated by the study in Ref.~\cite{Arina:2014yna} where the authors pointed out that the model with a light pseudoscalar mass $m_A\sim 50$ MeV and a $CP$-conserving coupling  between the Dirac DM and a pseudoscalar can provide good fits not only to the DAMA signal, but also to the correct relic density and GC gamma-ray excess.  However,  in Ref. \cite{Arina:2014yna},  the pseudoscalar propagator  squared $g_{p,\chi} g_p/(|\vec{q}|^2 + m_A^2)$ is replaced by the contact form $1/\Lambda^2$ in the direct-detection study (where the coupling $g_p$ will be defined in Sec. \ref{sec:DM-model}). We will show that such a replacement is not suitable. We take into account the full interaction form, because the value of $m_A$ could be of the same order of magnitude as the typical momentum transfer at the direct-detection experiments. Our results show a different conclusion: for the Yukawa-like couplings, under the correct relic density requirement, the DAMA signal can be accommodated only within a narrow parameter region where the long-range interactions, instead of contact interactions, occur in the DM-iodine scatterings.

We then show that the direct-detection rates are roughly proportional to $c_p^2 F_{\Sigma''}^{(p,p)} + c_n^2 F_{\Sigma''}^{(n,n)} + 2 c_p c_n F_{\Sigma''}^{(p,n)}$, where $c_p$ and  $c_n$ are proton and neutron effective couplings, respectively, and $F_{\Sigma''}^{N,N'}$ are response form factors with $N,N'\equiv n,p$. Choosing a suitable set of the quark spin contents of the nucleon and $m_u/m_d$,
the exclusion limits set by  XENON100 \cite{XENON100}, SuperCDMS \cite{SuperCDMS}, COUPP \cite{COUPP}, PICASSO \cite{PICASSO}, and LUX \cite{LUX} can be highly suppressed due to the destructive interference among  terms containing different response form factors, such that the DAMA signal is easily reconciled with these null measurements. 

We further explore the DAMA-allowed region constrained by $B$ and $K$ decays involving the pseudoscalar in the reaction. Because we consider a simplified model, the {\it effective} couplings between the pseudoscalar and quarks should originate from a higher scale through integrating out heavy states. Therefore, the induced flavor-changing neutral currents (FCNCs) may arise at the one-loop level from diagrams with quarks and $W$ bosons, such that the parameters can be further constrained by $B$ and $K$ decays. We estimate the flavor constraints, assuming that the relevant new physics occurs at the scale of 1 TeV \cite{Dolan:2014ska}.  In addition, we discuss parameter bounds required by the thermal freeze-out and astrophysical observations related to the big bang nucleosynthesis (BBN) and DM self-interaction.

The layout of this work is as follows. In Sec. \ref{sec:DM-model}, we introduce a general form of a Lagrangian that describes the interactions of a pseudoscalar mediator with SM quarks and the DM particle. The methods used in the analysis of the direct detections are described in Sec. \ref{sec:DirectDetection}. The relevant formulas for the relic abundance are presented in Sec. \ref{sec:relic}. Then, in Sec. \ref{sec:results}, we show the numerical results for DAMA, the correct relic density, and the null measurements in direct searches. Section~\ref{sec:discussions}   contains the parameter constraints  from $B$ and $K$ decays,  the requirement of the thermal freeze-out, and astrophysical observations related to the big bang nucleosynthesis and DM self-interaction. 
A summary is given in Sec.~\ref{sec:summary}.

\section{The Fermionic Dark Matter Model}\label{sec:DM-model}

We  focus on the study of the fermionic dark matter that couples to SM quarks via  a  pseudoscalar mediator $A$. For simplicity, we consider the DM particle to be a Dirac fermion, but the generalization to the Majorana fermionic DM case is straightforward. For the Dirac DM, we assume its chemical potential is negligible, i.e.,  the DM particle $\chi$ and antiparticle $\bar{\chi}$ have the equal number density. We include the $CP$-violating coupling between the DM and the pseudoscalar $A$. The effective Lagrangian is
 \begin{eqnarray}\label{eq:lagrangian-1}
{\cal L}_{\rm int}= A \overline{\chi}  
(g_{s,\chi}+i  g_{p,\chi}\gamma^5 ) \chi +  i A \sum_q g_p^q \bar q  \gamma^5 q \,,
\end{eqnarray} 
where $g_{p,\chi}$ and $g_{s,\chi}$ are the $CP$-conserving and $CP$-violating couplings, respectively.  For the Majorana fermionic DM, the factor 1/2 needs to be inserted in front of every Majorana fermionic bilinear; therefore, the expressions for direct detection and annihilation cross sections are identical with the Dirac case.  When we compute the scattering rate at direct-detection experiments, the interaction occurs with the whole nucleus scattered due to the small kinetic energy of the WIMP.  Therefore, we need to perform the calculation  at the nucleon level  with the Lagrangian rewritten as 
 \begin{eqnarray}\label{eq:lagrangian-2}
{\cal L}_{\rm int}= A \overline{\chi}  
(g_{s,\chi}+i  g_{p,\chi}\gamma^5 ) \chi +  i A \sum_{N=p,n}  c_N \bar N  \gamma^5 N \,,
\end{eqnarray}  
and further take into account nuclear form factors that describe the composite structure of the nucleus.
Here the effective coupling constants are given by
\begin{equation}
c_N=\sum_{q=u,d,s}  \frac{m_N}{m_q}  \left( g_p^q  - \sum_{q'=u,\dots,t} g_p^{q'} \frac{\bar m}{m_{q'}} \right) \Delta q^{(N)}
\end{equation}
where $\bar{m} =(1/m_u +1/m_d +1/m_s)^{-1}$ and the values
\begin{eqnarray}\label{eq:QuarkSpin}
\Delta u^{(p)} = \Delta d^{(n)} =+0.84 \,,\qquad
\Delta d^{(p)} = \Delta u^{(n)} =-0.44 \,,\qquad
\Delta s^{(p)} = \Delta s^{(n)} =-0.03 \,.
\label{eq:Deltaq}
\end{eqnarray}
are adopted for  the quark spin contents of the nucleon \cite{Cheng:2012qr} which, depending on the axial-vector form factors, are obtained using $(g_A^3, g_A^8, g_A^0)= (1.2701, 0.46, 0.37)$.

We will consider three different types of quark-pseudoscalar interaction: (i) quark universal couplings  for which  $g_p^q=g_p$ are constant, i.e., independent of quark flavors;  (ii)  quark Yukawa-like couplings for which  $g_p^{q}=g_p \sqrt{2}m_q/v$ with $v=246.2$ GeV, the vacuum expectation value of the SM Higgs; and (iii) quark first-generation couplings for which  $g_p^u=g_p^d=g_p \neq  0$, and zero for the rest. However, for simplicity, we exclude the couplings of the pseudoscalar to lepton sectors. 

The type of quark Yukawa-like couplings is consistent with the minimal flavor violation (MFV) ansatz \cite{D'Ambrosio:2002ex}, and can be related to Higgs-portal or axion-portal DM models \cite{Freytsis:2009ct}. The interaction with quark universal couplings has a non-MFV structure and has been studied in Refs. \cite{Arina:2014yna,Dolan:2014ska,DelNobile:2015lxa}; it introduces a larger $|c_p/c_n|$ ratio such that the DAMA signal can be easily reconciled with null direct-detection experiments  \cite{Arina:2014yna,DelNobile:2015lxa}.\footnote{However, we find that for the Yukawa-like couplings, if we choose another set of the quark spin contents of the nucleon given in Eq. (\ref{eq:Deltaq-2}) and $m_u/m_d\simeq 0.58$, the DAMA signal is instead reconciled with null measurements by LUX, XENON100, and SuperCDMS; this is not the case for quark universal couplings.}  However, because the flavor constraints may provide  stronger exclusions in most parameter regions, we further consider the interaction with quark first-generation couplings, for which the relevant FCNC couplings are reduced at least  by 5 orders of magnitude compared to the case of quark universal couplings.

\section{Direct detection}\label{sec:DirectDetection}

In a direct-detection experiment, the differential recoil rate for DM-nucleus scattering can be expressed as
\begin{eqnarray}\label{eq:rate}
\frac{dR_T}{dE_R}=N_T\frac{\rho_\odot}{m_{\chi}}\int_{v_{\text{min}}(E_R)}  v f_{\oplus}(\vec{v},t)\frac{d\sigma_T}{dE_R}   d^3v ,
\end{eqnarray}
where $N^{}_T$ is the number of target nuclei per unit mass, $E_R$ is the recoil energy of the target nucleus, $\rho_\odot \simeq 0.3$ GeV/cm$^3$ is the local DM density near Earth \cite{Freese:2012xd},\footnote{The local DM density 0.3 GeV/cm$^3$ has usually been adopted in direct-detection studies. However, a value between $0.2 \sim 0.8$ GeV/cm$^3$ is still allowed. Because we are interested in taking comparisons among direct-detection experiments, the conclusion is thus independent of the value of the local density.} $d\sigma_T/d E_R$ is the DM differential cross section, and  $f_\oplus(\vec{v},t)$ is the DM velocity distribution in the Earth frame.  Here $v_{\min}=\sqrt{m_T E_R/2\mu^2}$ is the minimal DM velocity needed for an elastic scattering with a recoil energy $E_R$ to occur, with $m_T$ being the mass of the target nucleus and $\mu=m_\chi m_T/(m_\chi+ m_T)$ being the reduced mass of the DM-nucleus system.
$f_\oplus(\vec{v},t)$ can be obtained in terms of the velocity distribution in the Galactic frame, $\tilde f(\vec{v})$, as
\begin{eqnarray}
f_\oplus(\vec{v},t)=\tilde f(\vec{v}+\vec{v}_\oplus(t)), 
\end{eqnarray}
where $\vec{v}_\oplus$ is the relative velocity of the Earth with respect to the Galactic frame, and its magnitude is approximately equal to its component projecting in the Galactic plane,
\begin{equation}
v_\oplus(t) \simeq \left[v_\odot +u_E \cos\gamma \cos \left( 2\pi \frac{t-152.5\ \text{days}}{365.25\ \text{days}}\right)\right] \, \text{km/s} \,,
\end{equation}
with  $v_\odot\simeq 232~ \text{km/s}$  arising from the Galactic rotational motion and the Sun's peculiar velocity. The relative velocity between the Earth and Sun has a value $u_E\simeq 30~\text{km/s}$ and is inclined of an angle $\gamma\simeq 60^\circ$ with respect to the Galactic plane \cite{Belli:2002yt,Schoenrich:2009bx,Freese:2012xd}.
We simply assume the DM velocity distribution in the Galactic frame  to be an isotropic Maxwellian distribution with cutoff at the Galactic escape velocity $v_{\text{esc}} = 544$ km/s \cite{Smith:2006ym},
\begin{eqnarray}
\tilde f_{\rm MB}(\vec{v};v_0,v_{\text{esc}})=\frac{1}{N_E}e^{-v^2/v^2_0}\Theta(v_{\text{esc}}-v) ,
\end{eqnarray} 
with $N_E=\pi^{3/2}v_0^3(\text{erf}(z)-2z\, \text{exp}(-z^2)\pi^{-1/2})$, $z=v_{\text{esc}}/v_0$, and  $v_0 = 220$ km/s being the most probable velocity.

For the direct-detection searches, the  relevant effective operators for the $CP$-conserving (CPC) and $CP$-violating (CPV) interactions between Dirac DM and nuclei can be represented as
\begin{eqnarray}
{\cal O}^{ N}_{\rm CPC}  &=& (\bar{\chi} i\gamma_5 \chi ) \, (\bar{N} i\gamma_5  N) \,, \\
{\cal O}^{ N}_{\rm CPV}  &=& (\bar{\chi}  \chi )\,  (\bar{N} i\gamma_5  N) \,,
\end{eqnarray} 
respectively. To compute the scattering amplitudes, we must take into account the bound-state effects and then sum up the interaction amplitudes over all nucleons in the nucleus.  The nuclear response of these types, different from  the standard spin-independent and spin-dependent responses that are usually adopted, has been systematically studied in Refs. \cite{Fitzpatrick:2012ix,Anand:2013yka}. First, we express the corresponding nonrelativistic operators in terms of the nucleon matrix elements of operators,
\begin{eqnarray}
\langle \chi(p'), \, N(k')| {\cal O}^{ N}_{\rm CPC} |\chi(p),\, N(k) \rangle  &\to& \quad\quad\, \, \, 4{\cal O}_6 = 4 (\vec{q}\cdot \vec{S}_\chi) (\vec{q}\cdot \vec{S}_N)\,, \\
\langle \chi(p'), \, N(k')| {\cal O}^{ N}_{\rm CPV} |\chi(p),\, N(k) \rangle  &\to& -4m_\chi {\cal O} _{10}=  -4 m_\chi i \vec{q}\cdot\vec{S}_N\,,
\end{eqnarray} 
where the momentum transfer is $\vec{q}=\vec{p}^{\, \prime}-\vec{p}$, and $\vec{S}_N$ and $\vec{S_\chi}$ are the nucleon spin and DM spin operators, respectively. 
 The differential DM-target nucleus interaction cross section reads
 \begin{equation}\label{eq:sigmaT}
\frac{\mathrm{d} \sigma_T}{\mathrm{d} E_R} (v, E_R) = \frac{1}{32 \pi} \frac{1}{m_\chi^2 m_T} \frac{1}{v^2} \overline{| {\cal{M}}_T |^2} \,,
\end{equation}
where
\begin{eqnarray}
\overline{| {\cal{M}}_T |^2} &=& \frac{16 g_{p,\chi}^2  }{(|\vec{q}|^2 +m_A^2)^2}\frac{1}{2 j_\chi +1} \frac{1}{2j+1} 
 \sum_{\text{spin}}  \sum_{N,N'=p,n}   c_N c_{N'}  \left| \langle \chi', T' |  O_6  \chi^+ \chi^-  N^+ N^- |\chi, T \rangle\right|^2  \nonumber \\
 &=& \frac{|\vec{q}|^4  g_{p,\chi}^2}{(|\vec{q}|^2 + m_A^2)^2} \frac{m_T^2}{m_N^2 }
  \sum_{N,N'=p,n} c_N \, c_{N'} \, F_{\Sigma''}^{N,N'}
  \end{eqnarray}
for the $CP$-conserving interaction and 
\begin{eqnarray}
\overline{| {\cal{M}}_T |^2} &=& \frac{16 m_\chi^2  g_{s,\chi}^2  }{(|\vec{q}|^2 +m_A^2)^2}\frac{1}{2 j_\chi +1} \frac{1}{2j+1} 
 \sum_{\text{spin}}  \sum_{N,N'=p,n}   c_N c_{N'}  \left| \langle \chi', T' |  O_{10}  \chi^+ \chi^- N^+ N^- | \chi,T  \rangle \right|^2  \nonumber \\
 &=& \frac{4 |\vec{q}|^2 g_{s,\chi}^2}{(|\vec{q}|^2 + m_A^2)^2} \frac{m_\chi^2 m_T^2}{m_N^2 }
  \sum_{N,N'=p,n} c_N \, c_{N'} \, F_{\Sigma''}^{N,N'}
  \end{eqnarray}
for the $CP$-violating interaction, with $j$ and $j_\chi$ being the spins of the nucleus and DM  particle, respectively. Here the mass difference of the proton and neutron is neglected, $(\chi^+, N^+)$ and $(\chi^-,N^-)$ are nonrelativistic fields involving only creation and  annihilation fields, respectively, for the DM particle $\chi$ and  nucleon $N$, and $T$ denotes the target nucleus. The nuclear form factors $F_{\Sigma''}^{N,N'}$, of  which the explicit results  for various nuclei can be found in Refs. \cite{Fitzpatrick:2012ix,Anand:2013yka}, are functions of  the dimensionless variable $y = (|\vec{q}|b/2)^2$, where $b\simeq [41.467/(45A^{-1/3}-25A^{-2/3})]^{1/2}$ is the harmonic oscillator parameter, and $|\vec{q}|=(2m_T E_R)^{1/2}$.

Finally, the rates can be expressed as
\begin{equation}
\label{eq:Rate-1}
\frac{\mathrm{d} R_T}{\mathrm{d} E_R} =
N_T \frac{\rho_\odot}{m_\chi} \frac{1}{32 \pi } \frac{m_T}{m_\chi^2 m_N^2}\,  
 \frac{|\vec{q}|^4  g_{p,\chi}^2}{(|\vec{q}|^2 + m_A^2)^2} 
\sum_{N, N' = p, n} c_N  \, c_{N'} \, \mathcal{F}_{\Sigma''}^{(N, N')}(y, T,t) \ ,
\end{equation}
for the $CP$-conserving interaction, and
\begin{equation}
\label{eq:Rate-2}
\frac{\mathrm{d} R_T}{\mathrm{d} E_R} =
N_T \frac{\rho_\odot}{m_\chi} \frac{1}{8 \pi } \frac{m_T}{ m_N^2}\,  
 \frac{|\vec{q}|^2  g_{s,\chi}^2}{(|\vec{q}|^2 + m_A^2)^2} 
\sum_{N, N' = p, n} c_N  \, c_{N'} \, \mathcal{F}_{\Sigma''}^{(N, N')}(y, T,t) \ ,
\end{equation}
for the $CP$-violating interaction,
where
\begin{equation}
\mathcal{F}_{\Sigma''}^{(N, N')}(y, T,t) \equiv \int_{v_{\text{min}}(E_R)} \hspace{-.50cm} \mathrm{d} ^3 v \, \frac{1}{v} \, f_\oplus(\vec{v},t) \, F_{\Sigma''}^{(N, N')}(y, T) \ .
\end{equation}
Note that for the case of the Dirac DM, no additional factor of  2 should appear in Eqs. (\ref{eq:Rate-1}) and (\ref{eq:Rate-2}) to account for the interaction due to $\bar{\chi}$, because such an effect has already been included in $\rho_\odot = m_\chi (n_\chi + n_{\bar{\chi}})$.

The nuclear response form factors $F_{\Sigma''}^{N,N'}$  are relevant to, with respect to the momentum transfer $\vec{q}$, the longitudinal component of the nucleon spin. It is interesting to note that,  in $|\vec{q}| \to 0$ limit,  we can obtain the following approximation:
\begin{eqnarray}
  \sum_{N,N'=p,n} c_N \, c_{N'} \, F_{\Sigma''}^{N,N'}(0)
 & \approx &\frac{4}{|\vec{q}|^2}  \frac{m_N^2 }{m_T^2}   \frac{1}{2J+1} \sum_{\text{spins for\ } T, T'}
  \left| \sum_{N=p,n} c_N \langle  T' |\vec{q}\cdot\vec{S}_N |T \rangle\right|^2  \nonumber \\
 & \approx &\frac{4}{3}\frac{J+1}{J}\left( c_p \langle S_p \rangle + c_n \langle S_n \rangle \right)^2\,,
 \label{eq:FF-physics}
\end{eqnarray}
where the spin average is taken for the target nucleus, $J$ is the spin of the initial target nucleus, and $\langle S_{p(n)} \rangle \equiv \langle T|S_{p(n)} |T\rangle $ are the expectation values of the proton (or neutron) spin for the nuclear ground state. 
In the nuclear shell model calculation, the unpaired nucleon mainly contributes to $\langle S_N \rangle$ and $J$,  such that we can have the approximation $c_p \langle S_p \rangle + c_n \langle S_n \rangle \to c_{\text{odd}}\langle S_\text{odd}\rangle$, where the subscript ``odd" represents for the kind of the unpaired nucleon \cite{Lewin:1995rx}. 
Therefore, considering various nuclides  that are relevant to the direct-detection experiments analyzed in this paper, only those with ground-state spins  $\geq 1/2$ [$^{19}\text{F}(1/2)$, $^{23}\text{Na}(3/2)$, $^{73}\text{Ge}(9/2)$, $^{127}\text{I} (5/2)$, $^{129}\text{Xe}(1/2)$, and $^{131}\text{Xe}(3/2)$] contribute to $F_{\Sigma''}^{N,N'}$, the spin-dependent form factors, of which the values at $|\vec{q}|=0$ are summarized in Table \ref{tab:NuclearResponse} \cite{Fitzpatrick:2012ix,Anand:2013yka}.

\begin{table}
\begin{tabular}{|c c c c |  c c  c |}
\hline
 &  $Z$ &   NA(\%)  & $J$ &  $F^{(p,p)}_{\Sigma''}(0)$   &  $F^{(n,n)}_{\Sigma''}(0)$  &  $F^{(p,n)}_{\Sigma''}(0)$  \\ \hline
$^{19}$F    & $9$    & 100 & 1/2     &  0.903 &  0.00030  &  $-0.0166$ \\ \hline
$^{23}$Na  & $11$  &100 & 3/2      & $0.132$  & $0.00084$  &  $0.0105$  \\  \hline
$^{73}$Ge &  $32$ & 7.7 & 9/2      &  $0.00010$ & $0.368$   &  $0.0061$\\ \hline
$^{127}$I   &  $53$ & 100 & 5/2    &   $0.130$     & $0.0080$ &  $0.0323$\\ \hline
$^{129}$Xe & $54$ & 26.4 & 1/2  &    $0.00021$  & $0.247$   &   $0.0072$\\ \hline
$^{131}$Xe & $54$ & 21.2 & 3/2  &    $0.000058$ &  $0.0878$   &  $0.00226$\\
\hline
 \end{tabular}
 \caption{Values of form factors $F_{\Sigma''}^{N,N'}$  at $|\vec{q}| = 0$ for the nuclides with nonzero spin, where NA $\equiv$ natural abundance and $J\equiv$ the spin of the nucleus. } \label{tab:NuclearResponse}
\end{table}

To evaluate the proton and neutron couplings, we use the current quark masses for the light quarks in the $\overline{\text{MS}}$ subtraction scheme~\cite{PDG},
\begin{eqnarray}\label{eq:LightMass}
m_u=2.3_{-0.5}^{+0.7}~ \text{MeV}, \qquad 
m_d=4.8_{-0.3}^{+0.5}~ \text{MeV}, \qquad
m_s=95\pm5~\text{MeV},
 \end{eqnarray}
  corresponding to the scale 2 GeV,
with the following ratio constraints:
\begin{eqnarray}\label{eq:MassRatio}
m_u/m_d=0.48\pm 10, \qquad 
m_s/((m_u + m_d)/2)=27.5\pm 1.0 \,.
 \end{eqnarray}
 For the heavy quarks, we use the running quark masses in the $\overline{\text{MS}}$ scheme \cite{PDG},
 \begin{eqnarray}\label{eq:HeavyMass}
m_c=1.275\pm0.025~ \text{GeV}, \qquad 
m_b=4.18\pm 0.03~ \text{GeV}, \qquad
m_t=160_{-4}^{+5}~\text{GeV}.
 \end{eqnarray}
 All quark masses  will be consistently rescaled to $\mu=1$~GeV. Using the central values of quark masses, the effective coupling constants and their ratios are given by
\begin{equation}
\begin{aligned}
 & c_p=-0.359 g_p \,,   && c_n=0.022  g_p \,,  & & c_p/c_n=-16.4\,,  && \quad \text{for quark universal couplings}\,, \\
 & c_p=-0.0115 g_p \,, && c_n=0.0028  g_p \,,& & c_p/c_n= -4.09\,, && \quad \text{for quark Yukawa-like couplings}\,, \\
 & c_p=3.18 g_p \,,      && c_n=-0.19  g_p \,,   & & c_p/c_n= -16.8\,, && \quad \text{for quark first-generation couplings}\,.
\label{eq:cN}
\end{aligned}
\end{equation} 

We fit the simplified model to the data using the Bayesian statistics. The approach is briefly described below.
\vskip0.3cm
\noindent{{\underline{\bf DAMA}}:}

The DAMA experiment, using highly radiopure NaI(Tl) scintillators, has observed an annual modulation in the energy spectrum  of the target sodium and iodine nuclei.  In the experiment, the measurable scintillation energy $E_\text{ee}$ (in electron-equivalent units, keVee) that is transferred from  the nuclear recoil energy $E_R$  can be written as$E_{\text{ee}} = q E_R$, where $q$ is called the quenching factor. In this paper, we will use the quenching factors  $q_{\text{Na}}\approx {0.3}$ and $q_I\approx0.09$ \cite{Bernabei:1996vj} for the sodium and iodine, respectively. To fit the annual modulation signal at DAMA to the theoretical models, we shall use the data points (in the first 12 energy bins) in the low-energy window ($2-8$) keVee, reported in Fig.~8 of Ref.~\cite{Bernabei:2013xsa}. We neglect the data points with energies larger than 8 keVee, because they do not show any statistically significant modulation.  
The $\chi^2$ is then given by \cite{Catena:2014uqa}
\begin{equation}
\chi^2_{\rm DAMA} = \sum_{i=1}^{12}\frac{1}{\sigma_{i}^{2}}\left[S_{\rm m}(\hat{E}^{i}_{\rm{ee}})-\hat{S}_{\rm m}(\hat{E}^{i}_{\rm{ee}})\right]^2 ,
\end{equation}
where $\sigma_i$ are the errors associated with the data points $\hat{S}_{\rm m}(\hat{E}^{i}_{\rm{ee}})$, and the expected annual modulation rate is averaged over the energy bin interval,
\begin{equation}
S_{\rm m}(\hat{E}^i_{\rm{ee}}) = \frac{1}{2\Delta \hat{E}_{\rm{ee}}} \int_{\hat{E}^i_{\rm{ee}}}^{\hat{E}^i_{\rm{ee}}+\Delta\hat{E}_{\rm{ee}}}
d\hat{E}_{\rm{ee}} 
\left(\frac{d R_T}{d\hat{E}_{\rm{ee}}}\bigg|_{\rm (June\ 2)} - \frac{d R_T}{d\hat{E}_{\rm{ee}}}\bigg|_{\rm (December\ 2)}\right) \,,
\end{equation}
with $\Delta \hat{E}_{\rm{ee}}=0.5$ keVee being the width of the energy bins.
The observable differential rate as a function of the scintillation energy can be represented by the convolution of the Gaussian energy resolution function and  potentially possible rate,
\begin{equation}
\label{rate_obs}
\frac{d R_T}{d\hat{E}_{\rm{ee}}}=   \int_{0}^{\infty} dE_{\rm{ee}} (2\pi\sigma^2)^{-1/2} \exp\left[-\frac{(E_{\rm{ee}}-\hat{E}_{\rm{ee}})^2}{2\sigma^2}\right]\frac{\partial E_{R}}{\partial E_{\rm{ee}}} \times \left( \frac{d R_T}{dE_{R}}\right)_{E_{R}=E_{R}(E_{\rm{ee}})} \,,
\end{equation}   
where  $\hat{E}_{\rm ee}$ is the actually observed energy,  $E_{\rm ee}$ is the energy potentially measurable, and the DAMA detector resolution is \cite{Farina:2011pw}
\begin{equation}
\sigma(E_{\rm{ee}}) = 0.448 \sqrt{E_{\rm{ee}}/{\rm keVee} }  +9.1\times 10^{-3} E_{\rm{ee}}/{\rm keVee} \,.
\end{equation}

\vskip0.5cm
\noindent{\underline{\bf Null direct-detection experiments}:}

Following the approach given in Ref. \cite{DelNobile:2013sia},  we determine the 90\% confidence level (CL) for exclusion limits from COUPP, PICASSO, SuperCDMS, XENON100, LUX. The analysis is based on the explicit formalism \cite{DelNobile:2013sia}, 
\begin{equation}\label{eq:DeltaChi2}
{\rm \chi^2_{\text{CL}}} (\lambda, m_\chi) = - 2 \sum_k N^\text{obs}_k \ln \left( \frac{N^\text{th}_k(\lambda,m_\chi) + N_k^{\rm bkg}}{N_k^{\rm bkg}} \right) + 2 \sum_k N^\text{th}_k(\lambda,m_\chi) \,,
\end{equation}
where $N^\text{obs}_k$,  $N_k^{\rm bkg}$,  and $N^\text{th}_k$ are the event numbers for the observation, expected background, and  theoretical prediction, respectively. Here each module or energy bin is denoted by $k$.

The $\chi^2$ value is chosen to be a certain CL, which corresponds to the bounds on the parameter $\lambda$. Here, to study the exclusion limits at 90\% CL,  we will  adopt $\chi_{\rm CL}^2=2.71$ for one single parameter $\lambda\equiv (g_{p,\chi} g_p)^{1/2} /m_A$ or $(g_{s,\chi} g_p)^{1/2} /m_A$, corresponding to the $CP$-conserving interaction  or $CP$-violating interaction.

 \section{Relic abundance constraints}\label{sec:relic}
 
We focus on the case of $m_A<m_\chi$,  which usually corresponds to $T_f^A < T_f^\chi$, where $T_f^A$ and $T_f^\chi$ are the freeze-out temperatures for $A$ and $\chi$, respectively.
 The Boltzmann equation for the DM $\chi$ of number density $n_\chi$ is given by
 \begin{equation}
 a^{-3} \frac{d (n_\chi a^3)}{dt}  =
\langle \sigma v_{\text{M\o l}}\rangle  \left[ (n_\chi^{(0)})^2 - n_\chi^2   \right] \,,
\nonumber   \label{eq:boltzmann}
 \end{equation}
 where  $\langle \sigma v_{\text{M\o l}}\rangle$  is the thermal average of $\sigma v_{\text{M\o l}}$, with $v_{\text{M\o l}}$ as the M\o ller velocity, and the equilibrium number density of the DM denoted as $n_\chi^{(0)}$. 
Well after the freeze-out temperature, the DM abundance is approximately constant within a comoving volume. By solving the Boltzmann equation,
 the present-day DM relic abundance and freeze-out temperature are given by \cite{Griest:1990kh,Gondolo:1990dk}
\begin{equation}
 \Omega_{\rm DM} h^2 \simeq  \eta\frac{1.04 \times 10^9\  {\rm GeV}^{-1}}{ J \sqrt{g_*} m_{\rm pl} }, \qquad
 x_f\simeq \ln \frac{0.0382 g_\chi m_{\rm pl} m_{\chi} \langle \sigma v_{\text{M\o l}}\rangle \delta(\delta+2) }{\sqrt{g_*  x_f}}, 
 \label{eq:xf}
  \end{equation}
where 
\begin{equation}
 J= \int_{x_f}^\infty \frac{\langle \sigma v_{\text{M\o l}}\rangle}{x^2} dx , 
\end{equation}
 $\eta$ = 2 (or 1) for the Dirac (or Majorana) DM particle, $h\simeq0.673$ is the scale factor for the present-day Hubble constant, $m_{\rm pl}\simeq 1.22\times10^{19}$ GeV is the Planck mass, $x_f \equiv m_\chi/T_f$, $\delta \equiv n_\chi (x_f)/n_\chi^{(0)}(x_f) - 1$ with $s$ the total entropy of the universe, $g_\chi=2$ is the number of degrees of freedom of the $\chi$ particle, and $g_*$ is the number of relativistic degrees of freedom (DOF). Here $g_*\gtrsim 87.25$, and we will adopt  $g_*\approx 87.25$, which is the sum of the relativistic DOF of the $A$ particle and SM for $4~{\rm GeV}<T_f <80~ {\rm GeV}$. The current value for the DM density, coming from global fits of
cosmological parameters to a variety of observations,  is $\Omega_{\rm{DM}}  h^2=0.1198 \pm 0.0026$ \cite{PDG,Ade:2013zuv}, which follows  $ x_f  J \sim \eta \times 0.634$ pb\,c, corresponding to a typical magnitude about the electroweak interaction.  It turns out that a convenient choice for the best-fit result is $\delta(\delta+2)=(n+1)$, where $n=0$ corresponds to  the $s$-wave annihilation, $n=1$ for $p$-wave annihilation, and so on \cite{Griest:1990kh}. Numerically, we obtain that $x_f\approx 20$ (or 21) for $s$-wave dominated annihilations if $m_\chi\sim10$ GeV (or $\sim$ 40 GeV), and  $x_f\approx 19$ (or 20) for $p$-wave dominated annihilations if $m_\chi\sim10$ GeV (or $\sim$ 40 GeV). The requirement of the correct relic abundance can further result in the constraint on the magnitudes of $m_A$ and the interacting strength between the thermal DM and SM particles.

For temperature $T\lesssim 3m_\chi$, the thermal average can be written as a single-integral formula~\cite{Gondolo:1990dk},
\begin{equation}
\langle \sigma v_{\text{M\o l}} \rangle = \frac{1}{8m_\chi^4 T K_2^2 (m_\chi/T)} \int_{4m_\chi^2}^{\infty}
\sigma  \sqrt{s} (s-4m_\chi^2)K_1 (\sqrt{s}/T) ds,
\end{equation}
where $K_{1,2}$ are the modified Bessel functions and $s$ is the center-of-mass energy squared. Note that it has been shown that  $\langle \sigma v_{\text{M\o l}} \rangle$ taken in the cosmic comoving frame is equal to the result that is performed in the laboratory frame in which one of the incoming particles is treated to be rest, i.e., $\langle \sigma v_{\text{M\o l}} \rangle =\langle \sigma v_{\rm lab}\rangle^{\rm lab}$ \cite{Gondolo:1990dk}. In the calculation, we expand the annihilation cross sections in powers  of $\epsilon$, $\sigma v_{\rm lab} =a + b \epsilon +\cdots $, where $\epsilon= (s-4m_\chi^2)/(4m_\chi^2)$ is the kinetic energy per unit mass in the laboratory frame. We can then obtain the thermally averaged annihilation cross sections   
\begin{equation}
\langle \sigma v_{\text{M\o l}} \rangle =a + \frac{3}{2} \frac{b}{x} + {\cal O}(x^{-2}),
\end{equation}
with $x\equiv m_\chi /T$. The abundance of the DM is determined by the $s$-channel annihilation into a SM quark pair through the exchange of the pseudoscalar $A$, and
by $t$- and $u$-channel annihilations into two $A$'s.  For the $CP$-conserving interaction, the thermally averaged cross section for the Dirac DM particles ${\bar\chi} \chi$ annihilating into a ${\bar q}q$ pair, which is a $s$-wave process, is given by
\begin{eqnarray}
  \langle \sigma v_{\text{M\o l}} \rangle_{{\bar \chi} \chi \to \bar{q} q} 
&\simeq& \sum_q 
   \left\{
\frac{g_{p,\chi}^2 \, g_p^q{}^2 n_c m_{\chi }^2}{2 \pi 
   \left[\left(m_A^2-4 m_{\chi }^2\right)^2+m_A^2\Gamma_{\text{CPC}}^2    \right] }  \sqrt{1-\frac{m_q^2}{m_{\chi }^2}} 
  \right.
   \nonumber\\
 &+& 
  \left.
 \frac{3 g_{p,\chi}^2 \, g_p^q{}^2  n_c  
 \left[   16 m_{\chi }^4 \left(m_A^2-4 m_{\chi}^2\right) +
 m_q^2 \left(80 m_{\chi }^4-24 m_A^2 m_{\chi}^2+m_A^4+m_A^2 \Gamma_{\text{PP}}^2  \right)
 \right]  }
   {8 \pi   \left[\left(m_A^2-4 m_{\chi }^2\right)^2+ m_A^2 \Gamma_{\text{CPC}}^2  \right]^2
    \sqrt{1-m_q^2/m_{\chi }^2} \,\, x }  
   \right\}   \,, \nonumber  \\
 \end{eqnarray}
where $n_c=3$ is the number of the quark's colors, and the $s$-wave contribution starts at ${\cal O}(x^0)$ and involves  ${\cal O}(x^{-1})$, which gives about $15\%$ correction to the leading term. For the $CP$-violating interaction, which is the $p$-wave process, this is obtained by
\begin{eqnarray}
 \langle \sigma v_{\text{M\o l}} \rangle_{{\bar \chi} \chi \to \bar{q} q} 
&\simeq& 
\sum_{q} \frac{ 3 g_{s,\chi}^2 \, g_p^q{}^2  n_c} {4 \pi} \frac{m_\chi^2}{ (4m_\chi^2-m_A^2)^2 + m_A^2 \Gamma_{\text{CPV}}^2}  
 \sqrt{1-\frac{m_q^2}{m_\chi^2}} \, \, \frac{1}{x}.
 \end{eqnarray}
For the thermally averaged cross section into  two $A$'s, the results can be expressed as
 \begin{eqnarray}  
&&\langle \sigma v_{\text{M\o l}} \rangle_{{\bar \chi} \chi \to AA} 
\simeq
\frac{g_{p,\chi}^4} {4 \pi} \frac{ m_\chi (m_\chi^2-m_A^2)^{5/2}}{ (2m_\chi^2-m_A^2)^4 } \frac{1}{x} \,,
\end{eqnarray}
 for the $CP$-conserving interaction, and 
 \begin{eqnarray}  
&&\langle \sigma v_{\text{M\o l}} \rangle_{{\bar \chi} \chi \to AA} 
\simeq
\frac{g_{s,\chi}^4} {4 \pi} \frac{ m_\chi^2 (9 m_\chi^4 - 8 m_\chi^2 m_A^2 + 2m_A^4)}{ (2m_\chi^2-m_A^2)^4 }  \sqrt{1-\frac{m_A^2}{m_\chi^2}} \, \, \frac{1}{x} \,,
\end{eqnarray}
 for the $CP$-violating interaction. It is interesting to note that the DM annihilations into SM quarks are $s$-wave and $p$-wave processes for $CP$-conserving and  $CP$-violating interactions, respectively, while the annihilation to the pseudoscalars is $p$-wave process for both interactions. The total widths of the $A$ for the two types of interactions are the same, $ \Gamma_{\text{CPC}} = \Gamma_{\text{CPV}} =\Gamma$, because $A\to {\bar \chi} \chi$ is forbidden for $m_A<m_\chi$. In the present case $m_A<m_\chi$, the width of the pseudoscalar satisfies $m_A\Gamma/(4m_\chi^2 )\ll 1$,  so that it can be negligible in the calculation (see also Fig. 2 of Ref.~\cite{Dolan:2014ska} and the discussion therein).

\section{Results}\label{sec:results}

\subsection{DAMA and null direct-detection measurements}

\begin{figure}[t!]
\begin{center}
\includegraphics[width=0.35\textwidth]{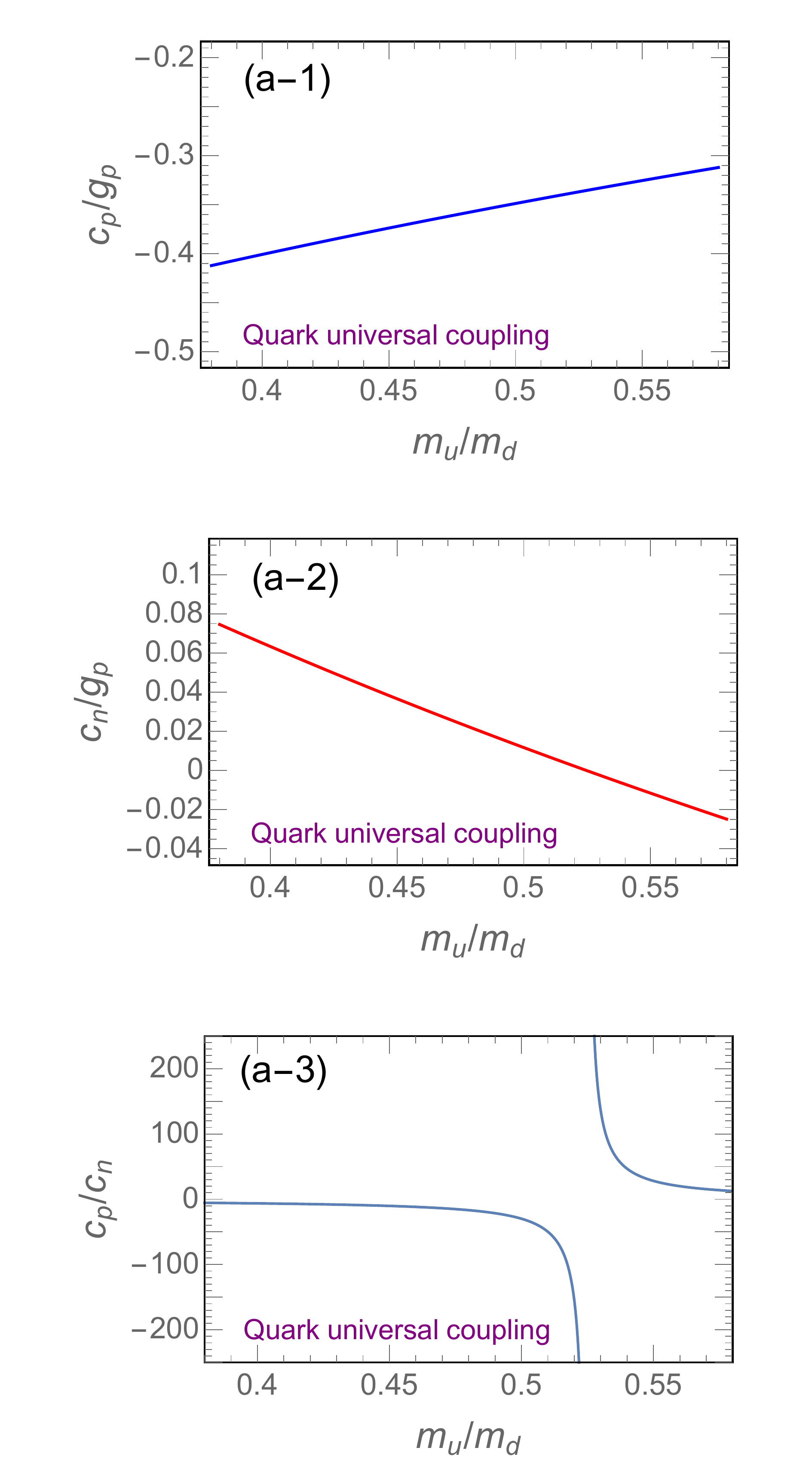}
\hspace{-0.65cm} \includegraphics[width=0.35\textwidth]{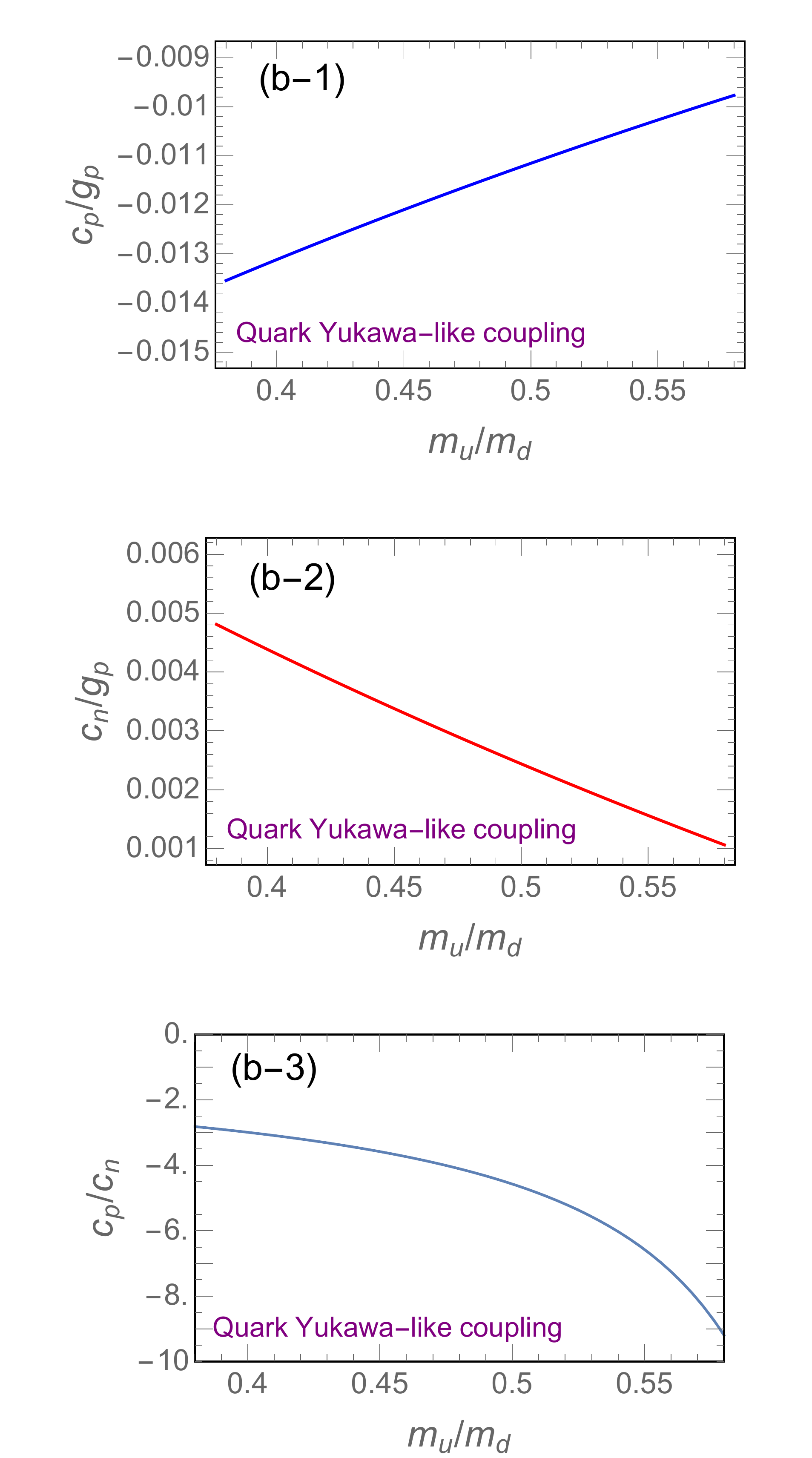}
\hspace{-0.65cm} \includegraphics[width=0.35\textwidth]{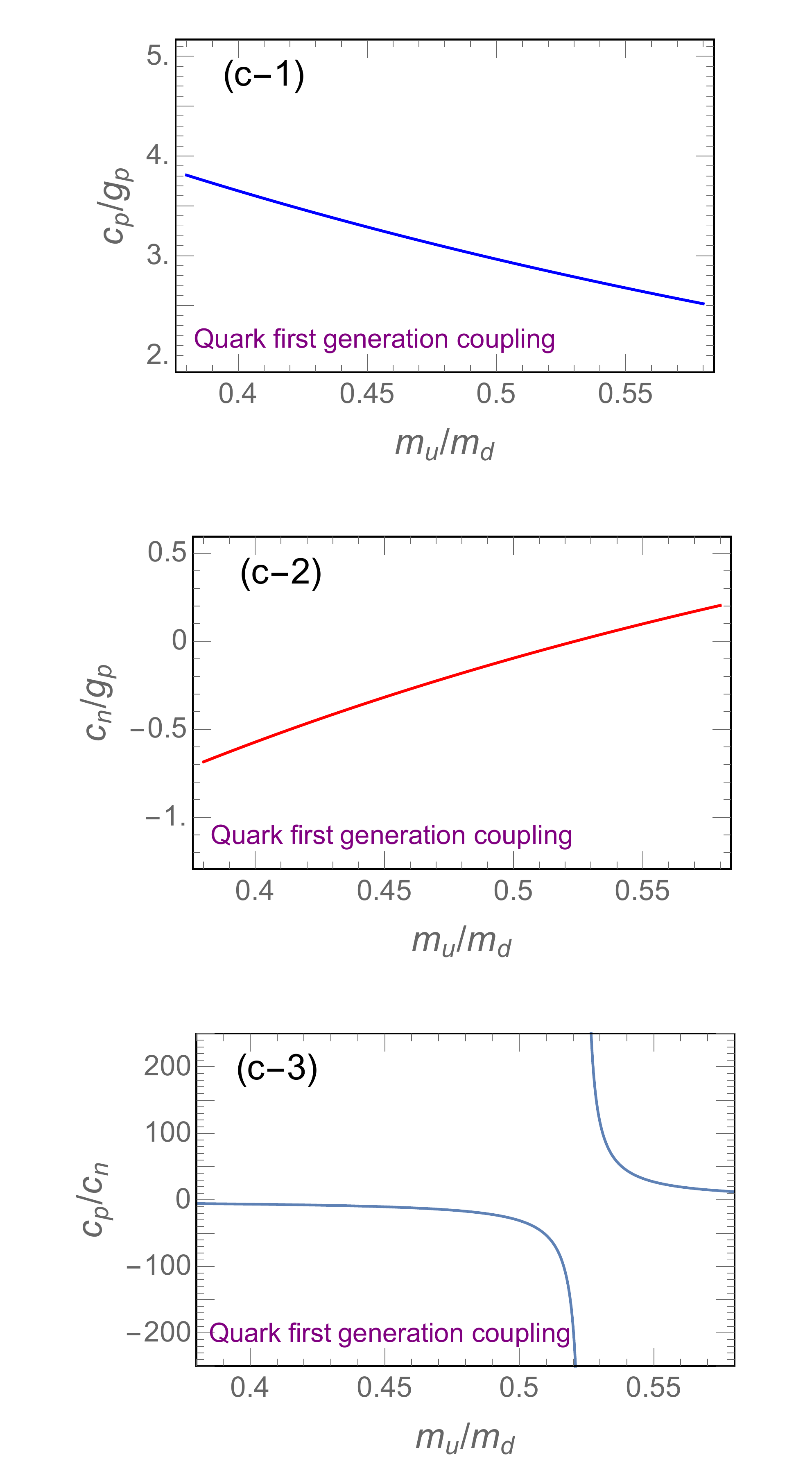}
\caption{The effective coupling constants as functions of $m_u/m_d$. The central values of quark masses ($m_u, m_s, m_c, m_b, m_t$) given in Eqs.~(\ref{eq:LightMass}) and (\ref{eq:HeavyMass}), and the values of  $\Delta q^{(N)}$'s  given in Eq. ( \ref{eq:QuarkSpin})  are used. }
\label{fig:MassRatio}
\end{center}
\end{figure}

\begin{figure}[t!]
\begin{center}
\includegraphics[width=0.96\textwidth]{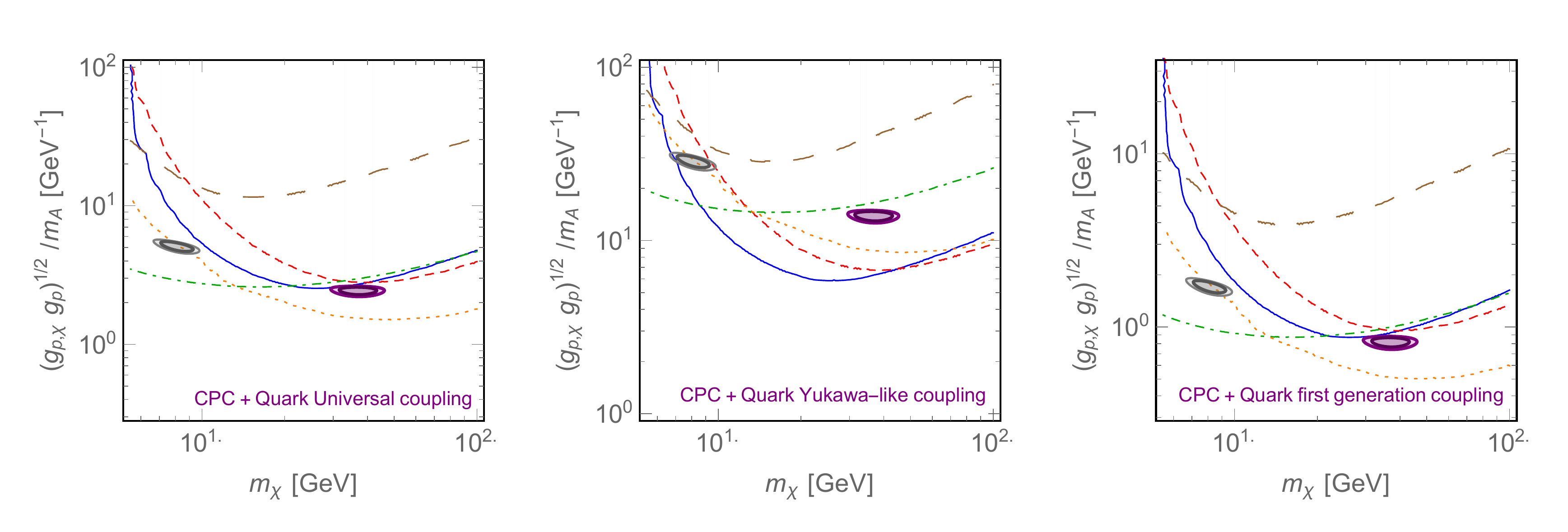}\\
\includegraphics[width=0.96\textwidth]{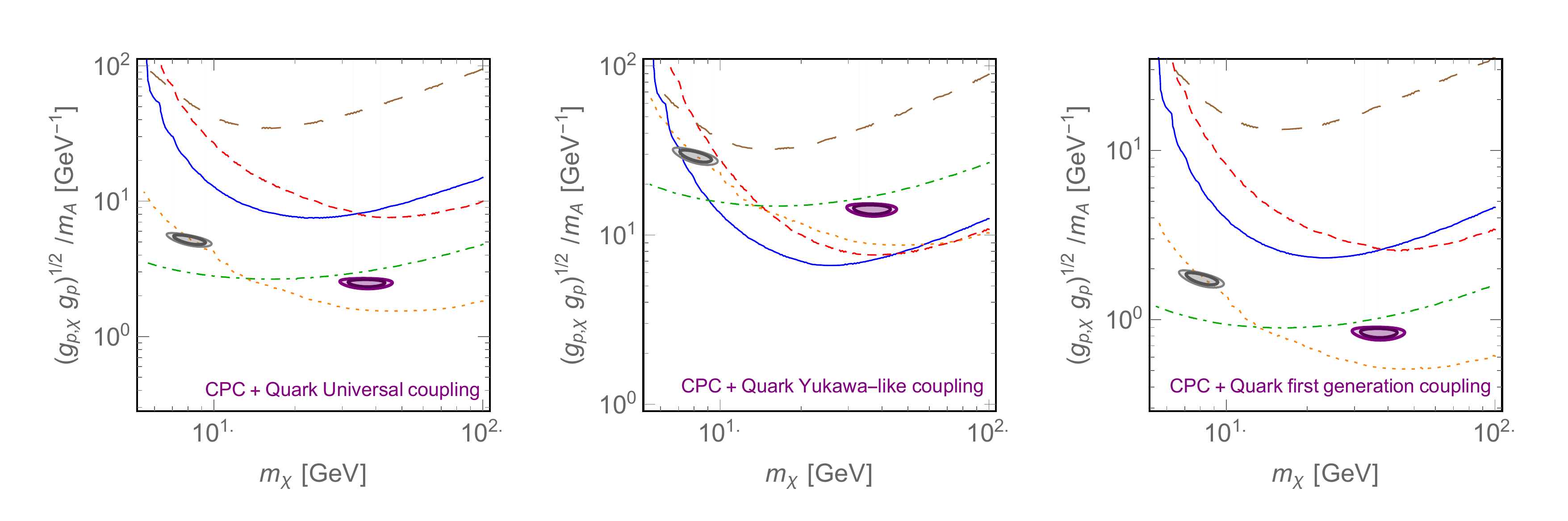}\\
\caption{The DAMA  2$\sigma$ (inner shaded region) and 3$\sigma$ (outer shaded region) allowed regions vs. 90\% CL upper limits from LUX (solid blue), XENON100 (dashed red), SuperCDMS (long-dashed brown), PICASSO (dot-dashed green), and COUPP (dotted orange) for the CPC, where we have taken $m_A=100$ MeV as a benchmark. For the DAMA, the two regions with  the gray color (at $m_\chi\sim$ 8 GeV)  and with the purple color (at $m_\chi\sim$ 37 GeV) correspond to scattering on the sodium and iodine, respectively. The central values of $\Delta q^{(N)}$ and quark masses (rescaled to $\mu=1$ GeV) given in Eqs. (\ref{eq:Deltaq}), (\ref{eq:LightMass}), and (\ref{eq:HeavyMass}) are used, except that $m_d=m_u/ 0.51$ is used in the lower panels.}
\label{fig:DirectDetectCPC}
\end{center}
\end{figure}

\begin{figure}[t]
\begin{center}
\includegraphics[width=0.98\textwidth]{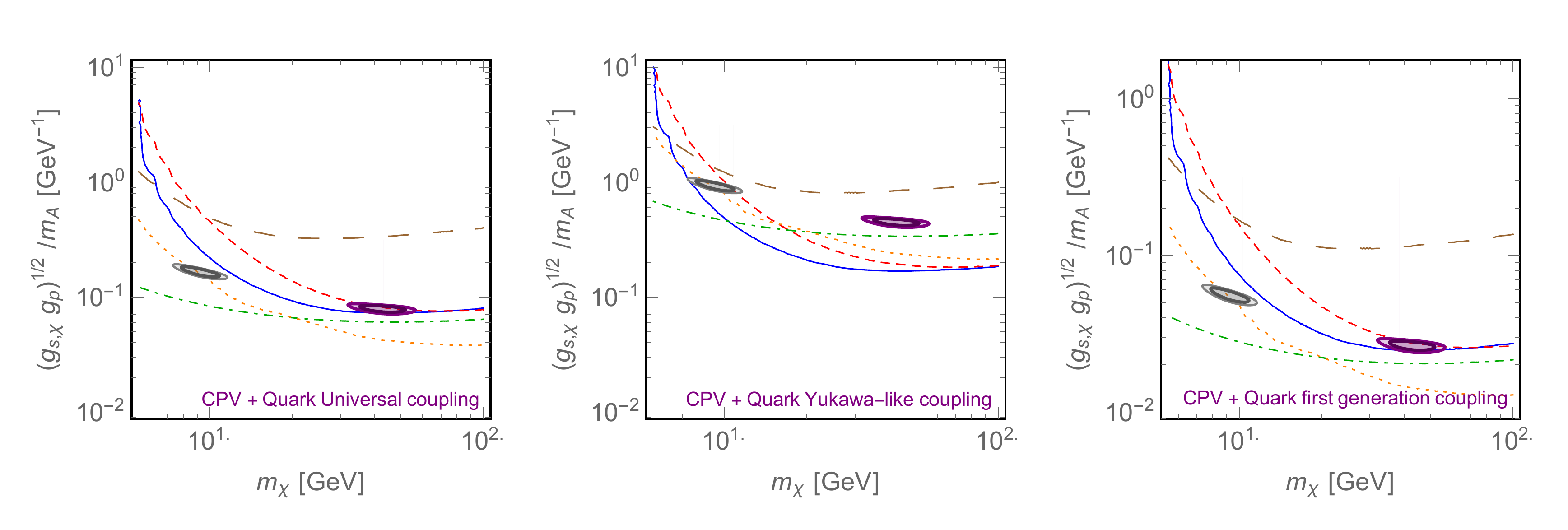}\\
\includegraphics[width=0.98\textwidth]{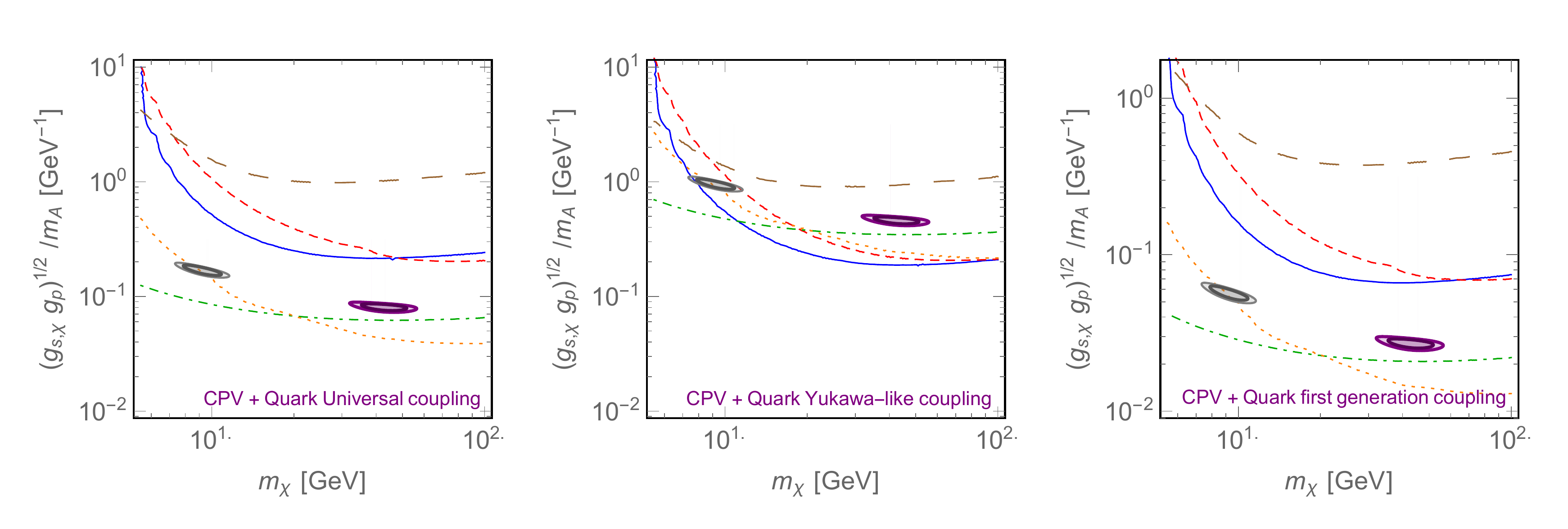}\\
\caption{Same as for Fig. \ref{fig:DirectDetectCPC} except that this is for  the CPV interaction.}
\label{fig:DirectDetectCPV}
\end{center}
\end{figure}

We fit the DAMA signal with two free parameters, $\frac{(g_{p (s),\chi} g_p )^{1/2}}{m_A}$ and $m_\chi$, with respect to various values of $m_A$. The results are summarized in Tables \ref{tab:CPC-U}$-$\ref{tab:CPV-F}.  The best fit is performed over 12-2 degrees of freedom. We obtain two qualitatively different best-fit regions, where one with a $m_\chi$ of 10 GeV is mainly due to scattering on  sodium and the other with $m_\chi $ of order 40 GeV is mainly due to scattering on  iodine.
Taking a good fit with a $p$ value $>0.05$ (95\% CL) which corresponds to $\chi^2 < 18.3$, we find that, for the $CP$-violating interaction, the fit with a heavier DM mass of order 40 GeV becomes poor when $m_A\lesssim$ 40 MeV. Here  $p=1/(2^{\nu/2} \Gamma(\nu/2))\int_\chi^\infty (C^2)^{(\nu-2)/2} \exp[-C^2/2]  d\text{C}$, with $\nu=12-2=10$, for which a too-low value of $p$ means the DAMA data are not consistent with being drawn from the model. For numerical results, the interesting points are as follows.
\begin{enumerate}

\item  Comparing Eq. (\ref{eq:Rate-1}) with Eq. (\ref{eq:Rate-2}), we know that, to fit the DAMA data, $g_{s,\chi}   \approx  g_{p,\chi} |\vec{q}|/(2m_\chi) \sim 10^{-3} g_{p,\chi}$ should be satisfied for the $CP$-violating interaction,  where the typical momentum transfer satisfies $|\vec{q}|\sim 80$~MeV if the signal is mainly due to scatterings on iodine or $|\vec{q}|\sim 20$~MeV  if it is mostly due to scatterings on sodium. Our results given in Tables \ref{tab:CPC-U}$-$\ref{tab:CPV-F} are consistent with this kinematic requirement that $g_{s,\chi}/g_{p,\chi}\simeq (1/30)^2$ under the same conditions for $m_A, g_p$ and $m_\chi$.

\item As shown in Tables \ref{tab:CPC-U}$-$\ref{tab:CPV-F},  for the best-fit solutions with $m_\chi$ of order 40 GeV and $m_A \gtrsim$ 300 MeV (or  with $m_\chi$ of order 10 GeV and $m_A \gtrsim$ 100 MeV), we can take the following replacement in the DAMA data analysis:
\begin{equation}
\label{eq:eff}
 \frac{g_{p(s),\chi}\,  g_p}{|\vec{q}|^2 + m_A^2} {\cal O}^N_{\rm CPC (CPV)}   \to   \frac{1}{\Lambda^2} {\cal O}^N_{\rm CPC (CPV)}\,,
   \end{equation}
where the resultant $\Lambda=m_A/ \sqrt{g_{p(s),\chi}\,  g_p}$ weakly depends on the value of $m_A$ owing to $m_A^2/(|\vec{q}|^2+m_A^2)\gtrsim0.95$. However, such a replacement is  invalid even when the value of $m_A$ is comparable with the typical momentum transfer $|\vec{q}|$; the deviation may give rise to different conclusions in the global analysis (see also Figs. \ref{fig:Relic-cpc} and \ref{fig:Relic-cpv}).

\item For $m_A \gtrsim$ 300 MeV, compared with the usual spin-independent case, the additional factors $|\vec{q}|^4$ and $|\vec{q}|^2$ for $CP$-conserving and $CP$-violating interactions, respectively (see also Eqs. (\ref{eq:Rate-1}) and (\ref{eq:Rate-2})), deplete the spectrum at low recoil energies and, hence, result in smaller DM masses for a good fit.
However, as for $m_A \lesssim |\vec{q}|$, the factor $(|\vec{q}|^2+m_A)^{-2} $ will further move the DM best-fit  regions to higher masses.

\item Because the detectors at experiments for DAMA, COUPP, PICASSO, SuperCDMS, XENON100, and LUX  \cite{XENON100,SuperCDMS,COUPP,PICASSO,LUX} are made of  NaI, ${\rm CF}_3{\rm I}$, ${\rm C}_4{\rm F}_{10}$, ${\rm Ge}$, ${\rm Xe}$, and  ${\rm Xe}$, respectively, the main contributions to DAMA, COUPP and PICASSO measurements arise from the unpaired protons (inside the abundant isotopes $^{19}\text{F}$, $^{23}\text{Na}$, or $^{127}\text{I}$), while the SuperCDMS, XENON100, and LUX data are mostly due to the contributions from the unpaired neutrons  (inside the abundant isotopes $^{73}\text{Ge}$, $^{129}\text{Xe}$, or $^{131}\text{Xe}$).\footnote{ Although $F_{\Sigma''}^{N,N'}$  for $ {\rm ^{12}C}$ are not given in Refs. \cite{Fitzpatrick:2012ix,Anand:2013yka}, their values equal to zero \cite{Catena:2015uha}. The results are expected since $ {\rm ^{12}C}$ has neither an unpaired proton nor an unpaired neutron.  In the present study, we neglect the contribution from ${\rm ^{13}C}$ due to the smallness of its natural abundance, 1.1\%.}

\item  In Fig. \ref{fig:MassRatio}, we show that as $m_u/m_d\approx 0.525$,  the value of $|c_p/c_n|$ goes to infinity (due to $c_n\to 0$) for interactions with quark universal couplings and quark first-generation couplings. Meanwhile, as  $m_u/m_d\approx 0.58$, a larger value for $|c_p/c_n|\approx 9$ can be obtained for the Yukawa-like couplings. The detection rates are roughly proportional to $c_p^2 F_{\Sigma''}^{(p,p)} + c_n^2 F_{\Sigma''}^{(n,n)} + 2 c_p c_n F_{\Sigma''}^{(p,n)}$. In experiments employing Xe (LUX and XENON100) and Ge (SuperCDMS) as detector materials, whose spins are mostly due to the unpaired neutron, the interference between $ 2 c_p c_n F_{\Sigma''}^{(p,n)}$ and $(c_p^2 F_{\Sigma''}^{(p,p)} + c_n^2 F_{\Sigma''}^{(n,n)})$  is destructive for $c_p/c_n<0$ (see also Table~\ref{tab:NuclearResponse}). 
 Numerically, we find that  as $c_p/c_n\approx -60\sim -40$,   the exclusion limits set by  SuperCDMS, XENON100, and LUX are highly suppressed\footnote{In the analysis, we use the LUX results published in 2014 \cite{LUX}. Though the LUX experiment has recently released the new result \cite{Akerib:2015rjg}, our conclusion remains unchanged.};  using Eq.~(\ref{eq:Deltaq}), we find that when $m_u/m_d \in (0.506,0.514)$ such suppression  occurs for the cases of quark universal couplings and quark first-generation couplings.
 The results are shown in the lower panels of Figs. \ref{fig:DirectDetectCPC} and \ref{fig:DirectDetectCPV}.\footnote{
All figures shown in this paper are relevant to the Dirac fermionic DM.  As for the Majorana DM, the allowed parameters can be approximately obtained with the following substitutions: 
$g_{p (s),\chi} \to g_{p (s),\chi}/2^{1/4}, \  g_p \to g_p/2^{1/4}$, and $m_A\to m_A/2^{1/4}$, which originate from the factor $\eta$ in Eq. (\ref{eq:boltzmann}).
}

\item One should note that $c_p/c_n$ depends not only on $m_u/m_d$, but also on the values of  $\Delta q^{(N)}$'s. Choosing $m_u/m_d=0.58$ and another set of $\Delta q^{(N)}$'s given in Ref.~\cite{Cheng:2012qr},
\begin{eqnarray}\label{eq:QuarkSpin-2}
\Delta u^{(p)} = \Delta d^{(n)} =+0.85 \,,\  \
\Delta d^{(p)} = \Delta u^{(n)} =-0.42 \,,\ \
\Delta s^{(p)} = \Delta s^{(n)} =-0.08 \,,
\label{eq:Deltaq-2}
\end{eqnarray}
corresponding to the use of $(g_A^3, g_A^8, g_A^0)= (1.2701, 0.585, 0.34)$, we find that  $c_p/c_n=-8.49, -8.30$, and $-22.7$ for the quark universal, quark first-generation, and quark Yukawa-like couplings, respectively, such that the Yukawa-like couplings can instead reconcile the DAMA signal with the null measurements of LUX, XENON100, and SuperCDMS. 
 
\item  Figures \ref{fig:DirectDetectCPC} and \ref{fig:DirectDetectCPV} show the contour plots for the DAMA modulation result and the upper bounds for  the null experiments in the $[m_\chi, (g_{p,\chi} g_p )^{1/2}/ m_A]$ and  $[m_\chi, (g_{s,\chi} g_p )^{1/2}/ m_A]$ planes, respectively, where $m_A=100$ MeV is used as a benchmark. We show the results using the central values of $\Delta q^{(N)}$ and quark masses given in Eqs.~(\ref{eq:Deltaq}), (\ref{eq:LightMass}) and (\ref{eq:HeavyMass})  as inputs. For comparison, we also show the plots using the same parameters except for $m_d=m_u/0.51$, which correspond to $c_p/c_n=-49.8$ and $-52.0$ for the cases of quark universal couplings and quark first-generation couplings, respectively, while $c_p/c_n=-4.9$ for the quark Yukawa-like couplings. For the DAMA, we present the best-fit region corresponding to 95.45\% (99.73\%) C.L. for two-dimensional parameter space, i.e., 2$\sigma$  (3$\sigma$) corresponding to   $\Delta\chi^2$= 6.18 ( $\Delta\chi^2$ = 11.83).
For the null experiments, we show the exclusion limits at 90\% CL.  The DAMA signal can be reconciled with the null measurements of LUX, XENON100, and SuperCDMS for  $c_p/c_n\approx -60\sim -40$. However, the DAMA signal is still in tension with COUPP and PICASSO because the target nuclei in these three experiments have unpaired protons.\footnote{A recent measurement \cite{Xu:2015wha} shows the Na quenching factor may be $\sim$ 0.19 at 6 keVee to $\sim$ 0.15 at 2 keVee, significantly smaller than that reported by the DAMA collaboration at low energies. A lower value of the quenching factor indicates larger recoil energies at DAMA and consequently favors a larger DM mass to fit the data. However, numerically we find that using a smaller (or larger) sodium or iodine quenching factor, encountered in the literature \cite{Xu:2015wha,Collar:2013gu}, does not significantly improve the fits.}  The COUPP and PICASSO experiments employing fluorine (F), which is light compared to sodium's mass, are relevant to constraining the DM scattering on sodium (Na) in the DAMA, while  COUPP also employing iodine (I)  can constrain the DM scattering on iodine in the DAMA.
\end{enumerate}

\begin{table}
\centering
\begin{tabular}{|c|cccc|cccc|} \hline 
\multicolumn{9}{|c|}{{$CP$-conserving interaction and quark universal couplings}}  \\ [0ex] \hline
\hline
$m_A$(GeV)& $\frac{(g_{p,\chi} g_p )^{1/2}}{m_A}$(GeV$^{-1}$) & $m_\chi$(GeV) & $\chi^2_{\rm min}$ & $p$ value & 
$\frac{(g_{p,\chi} g_p )^{1/2}}{m_A}$(GeV$^{-1}$) & $m_\chi$(GeV)& $\chi^2_{\rm min}$ & $p$ value \\
\hline
5.0     & $5.01$ & $8.07$ & $9.96$ & $0.44$    & $1.87$ & $33.2$ & $10.3$ & $0.41$ \\
1.0     & $5.01$ & $8.07$ & $9.96$ & $0.44$    & $1.87$ & $33.2$ & $10.3$ & $0.41$ \\
0.30   & $5.03$ & $8.08$ & $9.96$ & $0.44$    & $1.93$ & $33.8$ & $10.4$ & $0.41$ \\
0.10   & $5.11$ & $8.13$ & $10.0$ & $0.44$    & $2.45$ & $37.3$ & $10.8$ & $0.37$ \\
0.050 & $5.36$ & $8.36$ & $10.0$ & $0.44$    & $3.73$ & $41.0$ & $11.7$ & $0.30$ \\  
0.030 & $5.97$ & $8.74$ & $10.2$ & $0.42$    & $5.73$ & $42.7$ & $12.7$ & $0.24$ \\  
0.010 & $11.2$ & $10.2$ & $11.3$ & $0.35$       & $16.1$ & $42.6$ & $14.5$ & $0.15$ \\  
0.0020 & $51.4$ & $11.0$ & $11.8$ & $0.30$       & $79.1$ & $42.2$ & $15.0$ & $0.13$ \\  
\hline
\end{tabular}
\caption{ \small  Results of spectral  fits to the DAMA annual modulation signal with respect to various values of $m_A$, where the type of interaction is $CP$ conserving and  the pseudoscalar mediator couples to quarks with universal couplings. For $\chi_{\rm{min}}$, the corresponding $p$ value is given.
}
\label{tab:CPC-U}
\end{table}
 \begin{table}
\centering
\begin{tabular}{|c|cccc|cccc|} \hline 
\multicolumn{9}{|c|}{{$CP$-conserving interaction and quark Yukawa-like couplings}}  \\ [0ex] \hline
\hline
$m_A$(GeV)& $\frac{(g_{p,\chi} g_p )^{1/2}}{m_A}$(GeV$^{-1}$) & $m_\chi$(GeV) & $\chi^2_{\rm min}$ & $p$ value & 
$\frac{(g_{p,\chi} g_p )^{1/2}}{m_A}$(GeV$^{-1}$) & $m_\chi$(GeV)& $\chi^2_{\rm min}$ & $p$ value \\
\hline
5.0     & $28.2$ & $8.07$ & $9.96$ & $0.44$    & $10.6$ & $33.1$ & $10.3$ & $0.41$ \\
1.0     & $28.2$ & $8.07$ & $9.96$ & $0.44$    & $10.6$ & $33.1$ & $10.3$ & $0.41$ \\
0.30   & $28.2$ & $8.08$ & $9.96$ & $0.44$    & $11.0$ & $33.7$ & $10.4$ & $0.41$ \\
0.10   & $28.7$ & $8.15$ & $9.98$ & $0.44$    & $13.9$ & $37.0$ & $11.0$ & $0.36$ \\
0.050 & $30.2$ & $8.32$ & $10.1$ & $0.44$    & $21.2$ & $40.8$ & $11.8$ & $0.30$ \\  
0.030 & $33.5$ & $8.74$ & $10.2$ & $0.42$    & $32.4$ & $42.4$ & $12.8$ & $0.24$ \\  
0.010 & $63.1$ & $10.2$ & $11.1$ & $0.35$    & $90.8$ & $42.2$ & $14.7$ & $0.15$ \\  
0.0020 & $289$ & $10.9$ & $11.8$ & $0.30$     & $447$ & $41.8$ & $15.2$ & $0.13$ \\  
\hline
\end{tabular}
\caption{ \small Same as for Table \ref{tab:CPC-U} except for Yukawa-like couplings between the pseudoscalar mediator and quarks.}
\label{tab:CPC-Y} 
\end{table}
 \begin{table}
\centering
\begin{tabular}{|c|cccc|cccc|} \hline 
\multicolumn{9}{|c|}{{$CP$-conserving interaction and quark first generation couplings}}  \\ [0ex] \hline
\hline
$m_A$(GeV)& $\frac{(g_{p,\chi} g_p )^{1/2}}{m_A}$(GeV$^{-1}$) & $m_\chi$(GeV) & $\chi^2_{\rm min}$ & $p$ value & 
$\frac{(g_{p,\chi} g_p )^{1/2}}{m_A}$(GeV$^{-1}$) & $m_\chi$(GeV)& $\chi^2_{\rm min}$ & $p$ value \\
\hline
5.0     & $1.68$ & $8.07$ & $9.95$ & $0.44$    & $0.627$ & $33.2$ & $10.3$ & $0.41$ \\
1.0     & $1.68$ & $8.07$ & $9.95$ & $0.44$    & $0.629$ & $33.2$ & $10.3$ & $0.41$ \\
0.30   & $1.69$ & $8.08$ & $9.96$ & $0.44$    & $0.650$ & $33.8$ & $10.4$ & $0.41$ \\
0.10   & $1.71$ & $8.15$ & $9.97$ & $0.44$    & $0.822$ & $37.3$ & $10.8$ & $0.37$ \\
0.050 & $1.80$ & $8.30$ & $10.1$ & $0.44$    & $1.25$ & $41.1$ & $11.8$ & $0.30$ \\  
0.030 & $2.00$ & $8.74$ & $10.3$ & $0.42$    & $1.92$ & $42.7$ & $12.7$ & $0.24$ \\  
0.010 & $3.78$ & $10.2$ & $11.1$ & $0.35$       & $5.39$ & $42.6$ & $14.5$ & $0.15$ \\  
0.0020 & $17.3$ & $11.0$ & $11.8$ & $0.30$       & $26.6$ & $42.2$ & $15.0$ & $0.13$ \\  
\hline
\end{tabular}
\caption{ \small Same as for Table \ref{tab:CPC-U} except for that the pseudoscalar mediator directly couples only to first-generation  quarks with the universal couplings.}
 \label{tab:CPC-F} 
\end{table}
\begin{table}
\centering
\begin{tabular}{|c|cccc|cccc|} \hline 
\multicolumn{9}{|c|}{{$CP$-violating interaction and quark universal couplings}}  \\ [0ex] \hline
\hline
$m_A$(GeV)& $\frac{(g_{s,\chi} g_p )^{1/2}}{m_A}$(GeV$^{-1}$) & $m_\chi$(GeV) & $\chi^2_{\rm min}$ & $p$ value & 
$\frac{(g_{s,\chi} g_p )^{1/2}}{m_A}$(GeV$^{-1}$) & $m_\chi$(GeV)& $\chi^2_{\rm min}$ & $p$ value \\
\hline
5.0    & $0.163$ & $9.17$ & $10.8$ & $0.37$     & $0.0623$ & $37.5$ & $11.6$ & $0.31$ \\
1.0    & $0.163$ & $9.18$ & $10.8$ & $0.37$     & $0.0624$ & $37.6$ & $11.6$ & $0.31$ \\
0.30   & $0.163$ & $9.18$ & $10.8$ & $0.37$    & $0.0640$ & $38.6$ & $11.7$ & $0.31$ \\
0.10   & $0.166$ & $9.30$ & $10.8$ & $0.37$    & $0.079$ & $43.8$ & $12.4$ & $0.26$ \\
0.050 & $0.172$ & $9.65$ & $10.9$ & $0.37$    & $0.117$ & $47.7$ & $15.0$ & $0.13$ \\
0.030 & $0.189$ & $10.3$ & $11.0$ & $0.36$    & $0.171$ & $46.5$ & $21.1$ & $0.020$ \\  
0.010 & $0.345$ & $12.8$ & $12.0$ & $0.29$    & $--$ & $--$ & $--$ & $--$ \\  
0.0020 & $1.57$ & $14.1$ & $12.6$ & $0.25$    & $--$ & $--$ & $--$ & $--$ \\  
\hline
\end{tabular}
\caption{ \small  Same as for Table \ref{tab:CPC-U} except for the $CP$-violating interaction.}
\label{tab:CPV-U}
\end{table}

\begin{table}
\centering
\begin{tabular}{|c|cccc|cccc|} \hline 
\multicolumn{9}{|c|}{{$CP$-violating interaction and quark Yukawa-like couplings}}  \\ [0ex] \hline
\hline
$m_A$(GeV)& $\frac{(g_{s,\chi} g_p )^{1/2}}{m_A}$(GeV$^{-1}$) & $m_\chi$(GeV) & $\chi^2_{\rm min}$ & $p$ value & 
$\frac{(g_{s,\chi} g_p )^{1/2}}{m_A}$(GeV$^{-1}$) & $m_\chi$(GeV)& $\chi^2_{\rm min}$ & $p$ value \\
\hline
5.0    & $0.916$ & $9.17$ & $10.8$ & $0.37$     & $0.355$ & $37.3$ & $11.6$ & $0.31$ \\
1.0    & $0.916$ & $9.18$ & $10.8$ & $0.37$     & $0.356$ & $37.3$ & $11.6$ & $0.31$ \\
0.30  & $0.917$ & $9.18$ & $10.8$ & $0.37$    & $0.365$ & $38.3$ & $11.7$ & $0.31$ \\
0.10   & $0.930$ & $9.30$ & $10.8$ & $0.37$    & $0.450$ & $43.3$ & $12.5$ & $0.25$ \\
0.050 & $0.969$ & $9.65$ & $10.9$ & $0.37$    & $0.664$ & $47.2$ & $15.2$ & $0.11$ \\
0.030 & $1.06$ & $10.3$ & $11.0$ & $0.36$      & $0.967$ & $46.2$ & $21.7$ & $0.017$ \\  
0.010 & $1.94$ & $12.8$ & $12.0$ & $0.29$    & $--$ & $--$ & $--$ & $--$ \\  
0.0020 & $8.81$ & $14.1$ & $12.6$ & $0.25$    & $--$ & $--$ & $--$ & $--$ \\  
\hline
\end{tabular}
\caption{ \small  Same as for Table \ref{tab:CPC-U} except for the $CP$-violating interaction with Yukawa-like couplings between the pseudoscalar mediator and quarks.}
\label{tab:CPV-Y}
\end{table}

\begin{table}
\centering
\begin{tabular}{|c|cccc|cccc|} \hline 
\multicolumn{9}{|c|}{{$CP$-violating interaction and quark first generation couplings}}  \\ [0ex] \hline
\hline
$m_A$(GeV)& $\frac{(g_{s,\chi} g_p )^{1/2}}{m_A}$(GeV$^{-1}$) & $m_\chi$(GeV) & $\chi^2_{\rm min}$ & $p$ value & 
$\frac{(g_{s,\chi} g_p )^{1/2}}{m_A}$(GeV$^{-1}$) & $m_\chi$(GeV)& $\chi^2_{\rm min}$ & $p$ value \\
\hline
5.0    & $0.0547$ & $9.17$ & $10.8$ & $0.37$     & $0.0209$ & $37.5$ & $11.6$ & $0.31$ \\
1.0    & $0.0547$ & $9.17$ & $10.8$ & $0.37$     & $0.0210$ & $37.6$ & $11.6$ & $0.31$ \\
0.30   & $0.0548$ & $9.18$ & $10.8$ & $0.37$    & $0.0216$ & $38.6$ & $11.7$ & $0.31$ \\
0.10   & $0.0555$ & $9.30$ & $10.8$ & $0.37$    & $0.0266$ & $43.7$ & $12.4$ & $0.26$ \\
0.050 & $0.0579$ & $9.64$ & $10.9$ & $0.37$    & $0.0393$ & $47.8$ & $15.0$ & $0.13$ \\
0.030 & $0.0635$ & $10.2$ & $11.0$ & $0.36$    & $0.0574$ & $46.6$ & $21.1$ & $0.020$ \\  
0.010 & $0.116$ & $12.8$ & $12.0$ & $0.29$    & $--$ & $--$ & $--$ & $--$ \\  
0.0020 & $0.526$ & $14.1$ & $12.6$ & $0.25$    & $--$ & $--$ & $--$ & $--$ \\  
\hline
\end{tabular}
\caption{ \small  Same as for Table \ref{tab:CPC-U} except for the $CP$-violating interaction where the pseudoscalar mediator directly couples only to first-generation quarks with the universal couplings.}
\label{tab:CPV-F}
\end{table}

\subsection{Relic abundance constraint on the DAMA signal}

\begin{figure}[t]
\begin{center}
\includegraphics[width=0.95\textwidth]{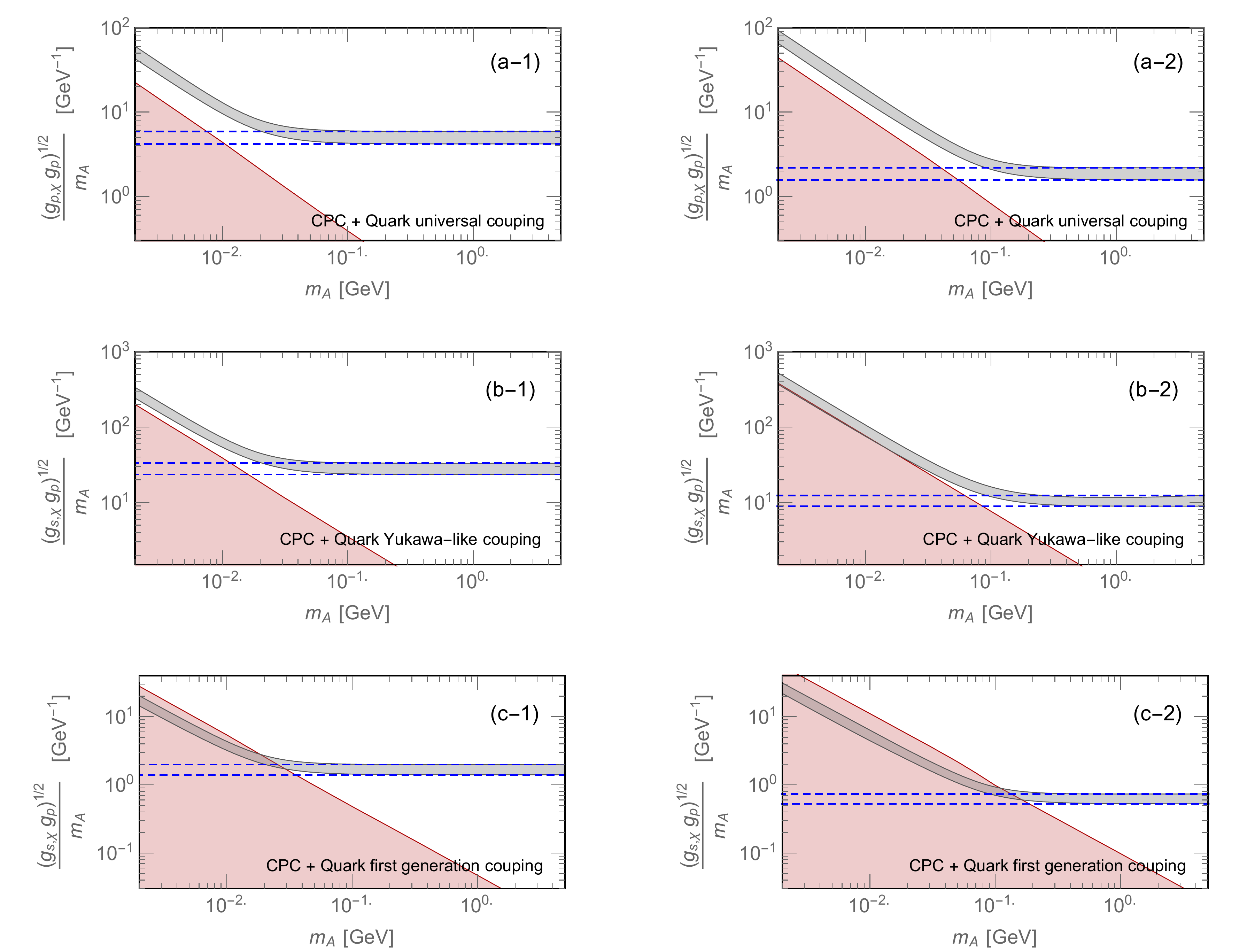}
\caption{   3$\sigma$-allowed DAMA regions (gray) vs. allowed regions for the correct relic abundance (red). The range between dashed lines is for the contact limit.
For the left panels, the DM mass is of order 10 GeV and the DAMA signal is dominated by scattering off sodium, while for the right panels the DM mass is of order 40 GeV and the DAMA signal is mostly due to scattering off the iodine target.}
\label{fig:Relic-cpc}
\end{center}
\end{figure}

\begin{figure}[t]
\begin{center}
\includegraphics[width=0.95\textwidth]{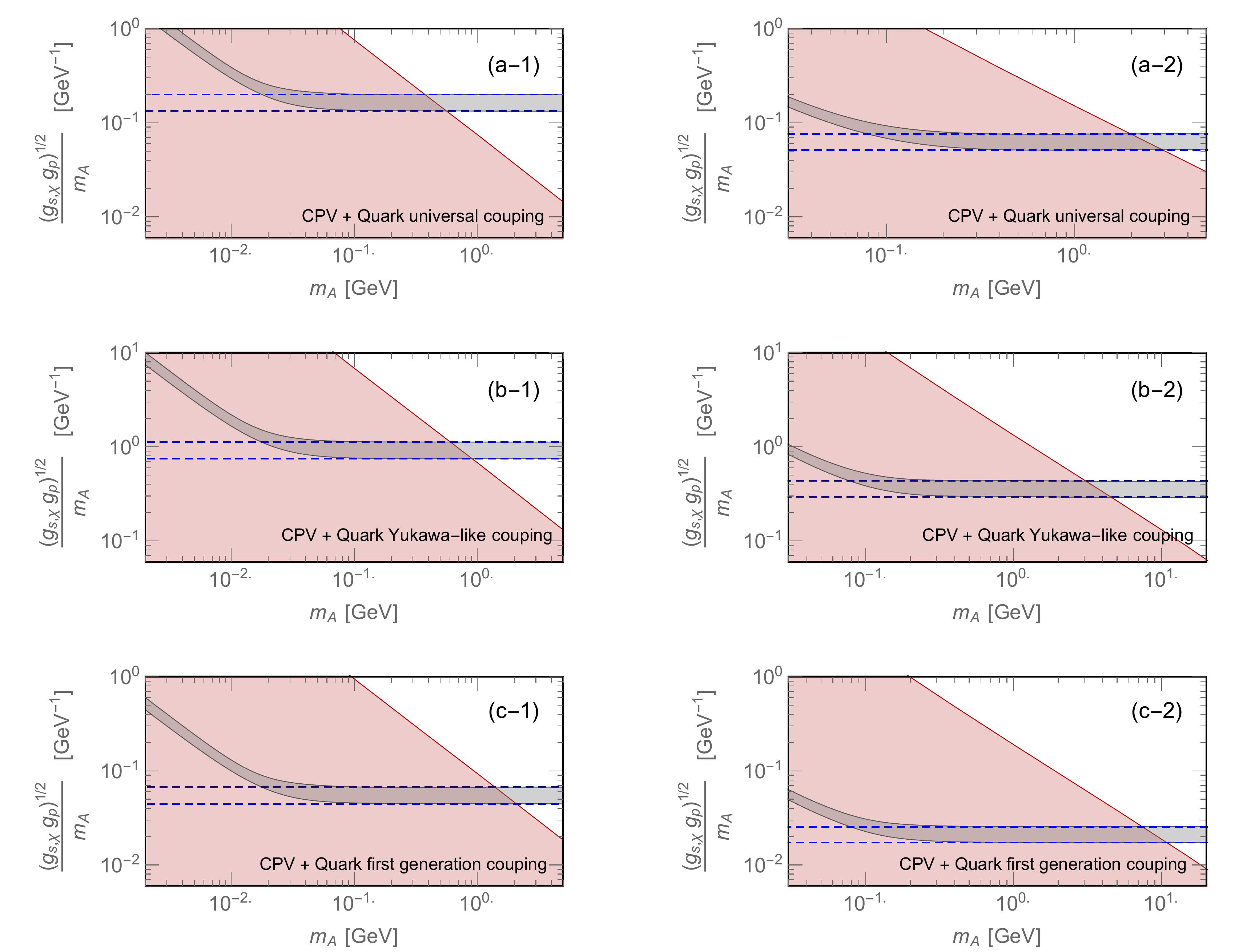}
\caption{Same as for Fig.~\ref{fig:Relic-cpc} except that this is for  the CPV interaction.}
\label{fig:Relic-cpv}
\end{center}
\end{figure}

The parameters extracted by the DAMA measurement can be further constrained by the requirement of  the correct relic density. We will substitute the value of $m_\chi$, obtained from the DAMA data, into the relic density  formula, i.e.,  $m_\chi$ is a function of $m_A$. In Figs.~\ref{fig:Relic-cpc} and \ref{fig:Relic-cpv}, we plot $3\sigma$-allowed DAMA regions, and also include uncertainties due to variations of quark masses and $\Delta{q}^{(N)}$, and correct relic density regions in the $(m_{A}, \sqrt{g_{p(s),\chi} g_p}/m_A)$ plane. 

The range between dashed lines corresponds to the DAMA signal extracted in the contact limit, with the replacement  ${g_{p(s),\chi} g_p}/({|\vec{q}|^2 +m_A^2}) \to 1/\Lambda^2$. Our results show that such a replacement is not suitable for the light mediator case.
Moreover, though the DAMA signal is incompatible with the correct relic density requirement for the $CP$-conserving interaction with quark universal couplings,\footnote{A similar conclusion for quark universal couplings was obtained in Ref. \cite{Dolan:2014ska}} for the $CP$-conserving interaction with Yukawa-like couplings only a small parameter region $m_A\lesssim$ 15 MeV can be accommodated, where the long-range interactions, instead of contact interactions, occur in the DM-iodine scatterings; the allowed result in Fig. 4(b2) is located in the DAMA iodine region with confidence level larger than 90\% ($\gtrsim 1.64\sigma$). (See Ref.~\cite{kcyPRLcomment} for further discussions on GC gamma-ray excess.)
Our results seem to indicate that if the pseudoscalar has universal couplings to quarks or Yukawa-like couplings to quarks, the fermionic DM-nucleus scattering mediated by a light pseudoscalar is dominated by the $CP$-violating interaction, i.e.,  $g_{s,\chi}  \gg 10^{-3} g_{p,\chi}$.

 It should be noted that although the case of quark first-generation couplings covers a widely allowed parameter region, it is, however, not helpful for explaining the GC gamma-ray excess. The result $m_\chi \approx 20$ GeV extracted from the relevant channel $\chi \chi \to {\bar q} q$ in the GC excess data  \cite{Daylan:2014rsa,Calore:2014nla},  is out of the range that we have obtained from the DAMA signal.

\section{Discussions}\label{sec:discussions}

\subsection{Flavor constraints}\label{sec:flavor}

The {\it effective} couplings between the pseudoscalar and quarks, which are not gauge invariant, should arise from a higher scale by integrating out heavy states. Interactions of these types may induce FCNCs, arising at the one-loop level from diagrams with quarks and $W$ bosons, such that the parameters can be further constrained by the (in)visible $B$ and $K$ decays. Here we will use the estimates given in  Ref.~\cite{Dolan:2014ska} where the relevant new physics is assumed to occur at the scale of 1 TeV. The resulting  Lagrangian for the induced FCNCs is given by
\begin{equation}
\mathcal{L}_\text{FCNC} = A \, \bar{d} (h^S_{ds} + h^P_{ds} \gamma^5) s + A \, \bar{s} (h^S_{sb} + h^P_{sb} \gamma^5) b + \text{H.c.}
\end{equation}
where the coefficients for the cases of quark-universal  and quark Yukawa-like couplings are,\footnote{For the Yukawa-like couplings, using the formula given in Ref.~\cite{Dolan:2014ska}, the value of $h^P_{sb}$ that we obtained is smaller by a factor of 3 compared to that in Ref.~\cite{Dolan:2014ska}. However, in the present study, flavor constraints are only relevant to  $h^S_{sb}$ and $h^S_{ds}$, not others.}
\begin{equation}
\begin{aligned}
h^S_{ds} & \approx \left(4.6 \cdot 10^{-6} + 2.0 \cdot 10^{-6} i \right) g_p \, , & h^P_{ds} & \approx \left(1.7 \cdot 10^{-6} + 7.3 \cdot 10^{-7} i\right) g_p\,, \\
h^S_{sb} & \approx \left(6.3 \cdot 10^{-4} - 1.2 \cdot 10^{-5} i \right) g_p \, , & h^P_{sb} & \approx \left(2.0 \cdot 10^{-4} - 3.8 \cdot 10^{-6} i \right) g_p \,,
\label{eq:h_u}
\end{aligned}
\end{equation}
and
\begin{equation}
\begin{aligned}
h^S_{ds} & = \left(3.5 \cdot 10^{-9} + 1.5 \cdot 10^{-9} i\right) g_p \, , & h^P_{ds} & = \left(3.9 \cdot 10^{-9} + 1.7\cdot 10^{-9} i\right)  g_p\,, \\
h^S_{sb} & = \left(2.3 \cdot 10^{-5} - 4.2 \cdot 10^{-7} i \right) g_p\, , & h^P_{sb} & = \left(2.3 \cdot 10^{-5} - 4.4 \cdot 10^{-7} i \right)  g_p\;  ,
\label{eq:h_y}
\end{aligned}
\end{equation}
respectively.
Following the calculations in Ref.~\cite{Dolan:2014ska}, the coefficients for the quark first-generation couplings are obtained as
\begin{align}
 & h^{S(P)}_{sb} = - \frac{\alpha \, g_p \, (m_b\pm m_s)  m_u}{16\pi \, m_W^2 \, \sin(\theta_W)^2} \, V_{ub} V_{us}^\ast \, \log\left(\frac{\Lambda^2}{m_t^2}\right) \; ,\\
 &  h^{S(P)}_{ds}  = - \frac{\alpha \, g_p \, (m_s\pm m_d)  m_u}{16\pi \, m_W^2 \, \sin(\theta_W)^2} \, V_{us} V_{ud}^\ast \, \log\left(\frac{\Lambda^2}{m_t^2}\right) \; ,
\end{align}
where we will adopt $\Lambda$=1 TeV corresponding to the new physics scale, and $m=m_t$, the relevant mass scale for the process under consideration.  Numerically, we obtain  
\begin{equation}
\begin{aligned}
h^S_{ds} & \approx \left(-2.5 \cdot 10^{-11}  \right)  g_p \, , & h^P_{ds} & \approx \left(-2.3 \cdot 10^{-11}  \right)  g_p, \\
h^S_{sb} & \approx \left(-1.8\cdot 10^{-12} + 4.5  \cdot 10^{-12} i\right) g_p \, , & h^P_{sb} & \approx \left(-1.8 \cdot 10^{-12} + 4.4\cdot 10^{-12} i\right)  g_p.  
\label{eq:h_1st}
\end{aligned}
\end{equation}

We have considered  the scenarios that $m_A<m_\chi$, and that dark matter couples only to the Standard Model quarks through the pseudoscalar particle exchange directly. Because $m_A<2 m_\chi$, the pseudoscalar $A$ has no invisible decay modes. Some possible experimental channels that can constrain the parameters relevant to the present models will be studied in this subsection. It is interesting to note that the $A$'s lifetime could be so long that it escapes the detector without decaying and, thus, behaves like an invisible particle  \cite{Dolan:2014ska}.  However, we do not consider this situation. 

\subsubsection{B-meson decays}

CLEO has reported an upper limit for the FCNC processes, including $b\to sg, dg, sq{\bar q}, dq{\bar q}$, which were referred to collectively as $b\to s g$~\cite{Coan:1997ye}.  In addition to using the fact  that the partial width $\Gamma(B \rightarrow X_s \, A)$ should be smaller than the total width $\Gamma_{B_s}^{\text{exp}}$, the experimental bound $\text{Br}^{\rm exp}(b\to s g)<6.8\%$ can be set as the upper limit for $\text{Br}(B \rightarrow X_s \, A)$, where $A$ decays hadronically and its corresponding mass should be larger at least than $3m_\pi$ due to $CP$ symmetry \cite{Dolan:2014ska}. The semi-inclusive decay width for $B \to X_s \, A$ is \cite{Hiller:2004ii}
 \begin{equation}
\Gamma(B \rightarrow X_s \, A) = \frac{1}{8 \pi} \frac{\left(m_{b}^2 - m_A^2\right)^2}{m_b^3} |h^S_{sb}|^2 \, ,
\label{eq:BXwidth}
 \end{equation}
where $X_s$ is an arbitrary hadron containing a strange quark.  The measurement of the inclusive partial branching ratio, 
\begin{equation}
\text{Br}^{\rm exp} (B^0 \to K^0 X) =(195^{+51}_{-45}\pm 50) \times 10^{-6} \,,
\end{equation}
reported by {\it BABAR} \cite{delAmoSanchez:2010gx} in the region where the momentum of the $K$ is greater than 2.34 GeV in the $B$ rest frame, can be used as the upper limit for $\text{Br}(B \rightarrow K \, A)$. Using the factorization approximation, the two-body decay width for $B \rightarrow K \, A$ is given by
\begin{equation}
\Gamma(B \rightarrow K \, A) = \frac{1}{8 \pi \, m_{B}^2} p_c(m_{B}^2, m_{K}^2, m_A^2) \left[ F_0^{BK}(m_A^2)\right]^2 \left(\frac{m_{B}^2 - m_{K}^2}{m_b - m_s}\right)^2 |h^S_{sb}|^2 \; ,
\label{eq:Bwidth}
\end{equation}
with $p_c(a,b,c) \equiv [(a^2 -b^2-c^2)^2-4\, b^2\, c^2 ]^{1/2}/(2a)$ and the light-cone sum rule result for the $B\to K$ form factor as \cite{Ball:2004ye}
\begin{equation}
F_0^{BK}(q^2) = \frac{0.330}{1- q^2/(37.46\ {\rm GeV}^2) } \; .
\end{equation}

\subsubsection{Kaon decays}

The $K_{\mu2}$ experiment at KEK has also measured  the momentum spectrum of the $\pi^+$  in the two-body $K^+$ decay \cite{Yamazaki:1984vg} to search for  a bump due to a neutral boson. The upper bound of  such a neutral boson covers the mass range from 10 to $300~{\rm MeV}$, except for the narrow range around the $\pi^0$ mass  where the limit becomes weaker.
This experimental result can be used to constrain $m_A$ and the coupling parameter $g_p$ from the $K^+ \rightarrow \pi^+ \, A$ decay, for which the decay width, in analogy to $B \rightarrow K \, A$, is
\begin{equation}
\Gamma(K^+ \rightarrow \pi^+ A) = \frac{ 1}{8 \pi \, m_{K^+}^2}
 p_c(m_{K^+}^2, m_{\pi^+}^2, m_A^2) 
 [F_0^{K\pi}(m_A^2)]^2
  \left(\frac{m_{K^+}^2 - m_{\pi^+}^2}{m_s - m_d}\right)^2 |h^S_{ds}|^2 \,,
\label{eq:Kwidth}
\end{equation}
with $F_0^{K^+}(m_A^2)\simeq 1$ \cite{Marciano:1996wy}. 
The experimental bound for the branching ratio was found  be to $\sim 10^{-6}$ at 90\% C.L. for the mass range $m_A \lesssim 70$ MeV, relaxing to $10^{-5}$ at $m_A \sim120$ MeV. For the numerical analysis, we take the bound from Fig. 2 of Ref. \cite{Yamazaki:1984vg}, together with the constraint obtained by requiring its branching ratio to be less than $\text{Br}^{\rm exp} (K^+ \to \pi^+ \pi^0) =(20.67\pm 0.08)\%$ \cite{PDG} for the region $m_A\sim m_\pi$.

\subsubsection{Results}

\begin{figure}[t]
\begin{center}
\includegraphics[width=0.33\textwidth]{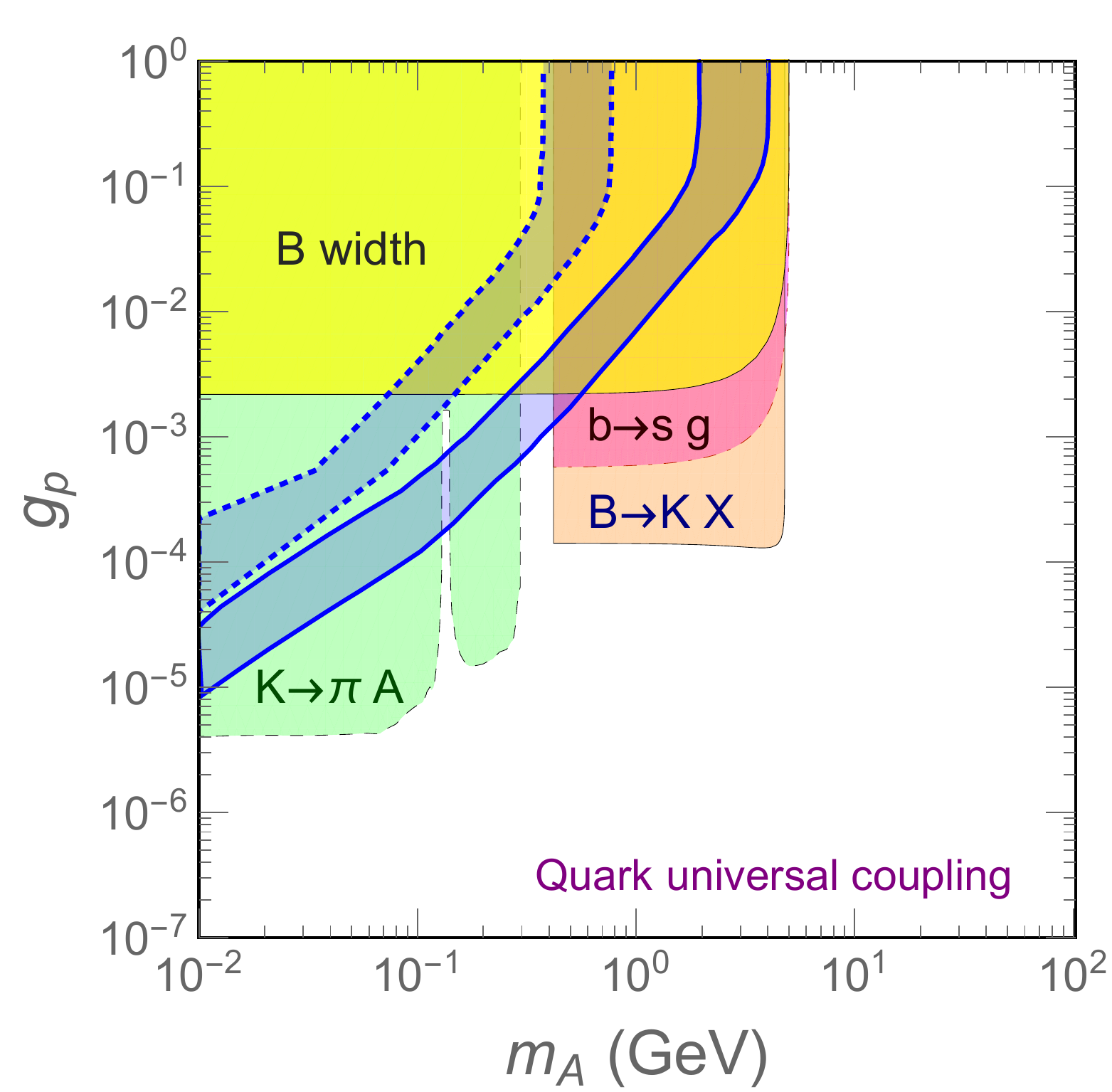}
\hspace{-0.2cm} \includegraphics[width=0.33\textwidth]{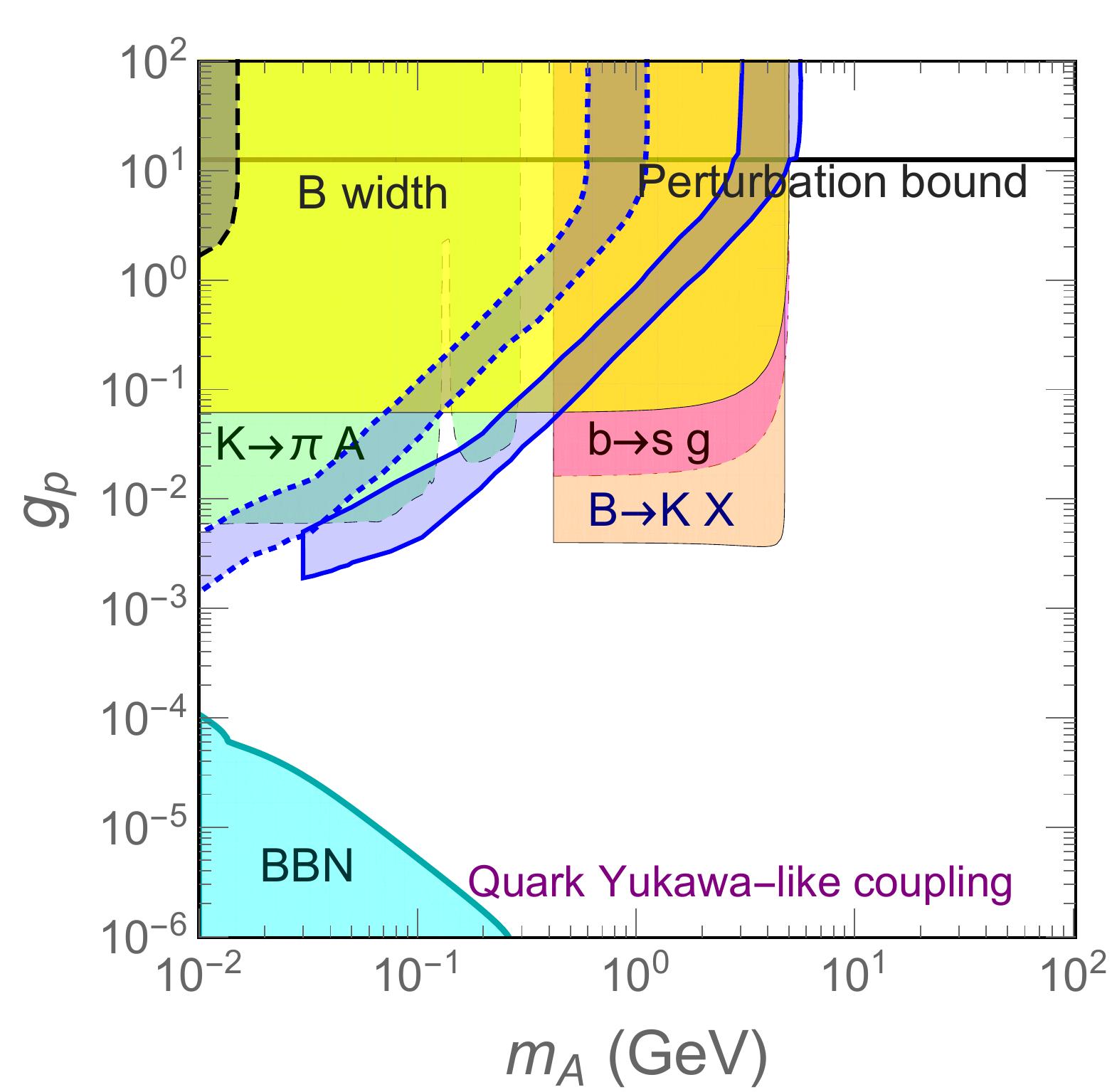}
\hspace{-0.2cm} \includegraphics[width=0.33\textwidth]{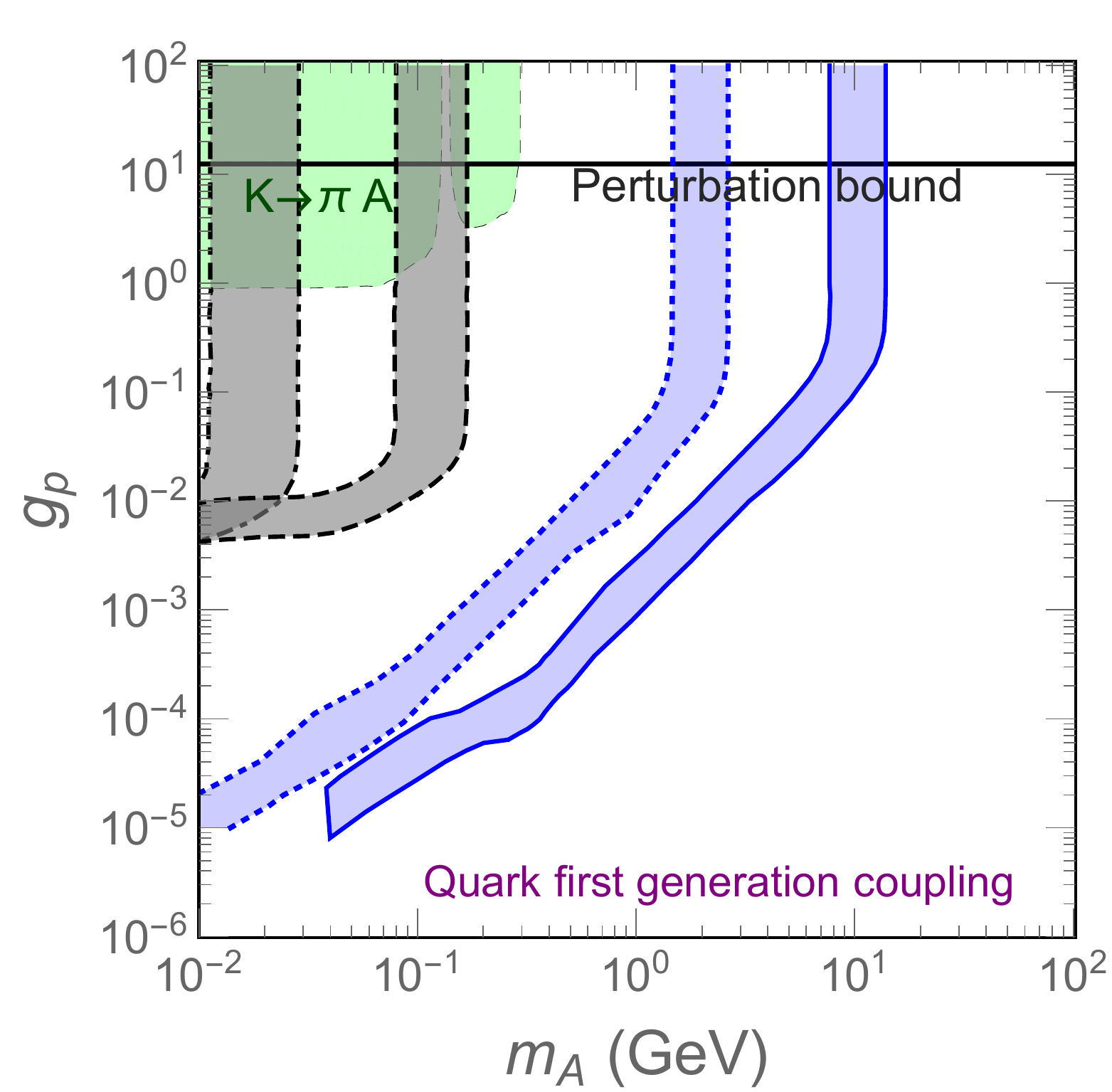}
\caption{ Allowed $(m_A, g_p)$ parameter regions determined by the relic abundance constraint and DAMA data, where the gray regions are for the $CP$-conserving interaction and blue ones for the $CP$-violating interaction. The DAMA regions for the DM particle scattering on Na are bounded by  the dot-dashed and dotted lines (where $m_\chi \sim10$~GeV),  and on I  are bounded by the  dashed and solid boundaries  (where $m_\chi \sim40$~GeV). Also shown are the exclusion contours on the $(m_A, g_p)$ plane from the various precision measurements and BBN, where the excluded regions with colors of yellow, red, orange, green, and cyan are, respectively, related to the $B$ width, $b\to g$, $B\to K X$, $K\to \pi A$, and BBN. The horizontal line depicts the maximum value ($4\pi$) that allows the perturbative calculation to be valid.}
\label{fig:FlavorBound}
\end{center}
\end{figure}

In Fig. \ref{fig:FlavorBound}, we display the excluded regions set by various precision measurements of $B$ and $K$ decays and compare them with the allowed $(g_p,m_A)$ parameter regions determined by the DAMA signal and correct relic abundance.

For  quark universal couplings and quark Yukawa-like couplings, only  the $CP$-violating interaction allows small parameter regions, where the corresponding thermally averaged annihilation cross sections are dominated by $\langle \sigma v_{\text{M\o l}}\rangle_{\bar{\chi}\chi\to A A }$.  For the former, the allowed regions are close to $m_A\sim m_\pi$ and $\lesssim 3 m_\pi$, where the DM mass is of order 40 GeV. For the latter, the flavor constraints exclude almost all DAMA regions with  $m_\chi\sim 10$ GeV apart from a small triangle region for $m_A\lesssim 30$~MeV, while the DAMA regions with $m_\chi\sim 40$ GeV are excluded except for $30< m_A< 420$ MeV.
 Our results show that if the pseudoscalar couples only to $u$ and $d$ quarks with the same coupling, the flavor physics will provide a considerably weaker constraint due to the fact that the FCNC couplings $h_{ds}^S$ and $h_{sb}^S$ are reduced by about 5 and 8 orders of magnitude, respectively, compared to the case of quark universal couplings.

In summary, if the magnitudes of FCNC coefficients are not overestimated, the cases of quark universal couplings and quark Yukawa-like couplings are strongly constrained by $B$ and $K$ decays, and very narrow parameter regions are allowed. However, since  the simplified model  is a model-independent  bottom-up approach, a phenomenological extension  of this model may change the values of FCNCs and the resultant flavor bounds.

\subsection{Bounds from other requirements}\label{sec:others}

In the following, we will discuss some parameter constraints which may be required by the thermal freeze-out and astrophysical observations.
  
\noindent{\underline{\bf Thermal equilibrium between the DM and visible sectors}:}

First, let us briefly discuss the lower bound of the DM-quark coupling $g_p$ necessary for obtaining thermal equilibrium between the DM and visible sectors.
For a too-small value $g_p$, the thermally averaged annihilation cross section will be dominated by the process $\chi {\bar \chi} \to AA$, such that the DM may have a different thermal temperature compared with the visible sector due to the decouple of the interactions between them.

In order for the DM to still maintain the same temperature with the visible sector before freeze-out, we impose the condition that the reaction rate is larger than the expansion rate of the Universe: $\sum_q \langle \sigma v_{\text{M\o l}}\rangle_{\bar{q} q \to \bar{\chi}\chi} n_{\rm eq}^q \gtrsim H$, where $n_{\rm eq}^q$ is the thermal number density of the quark {\it q}, and the left-hand side is expected to equal to the production rate of SM particles from the DM annihilation in the thermal equilibrium \cite{Dolan:2014ska}. The results for the $CP$-violating cases can be obtained in the same manner as the $CP$-conserving ones as given in \cite{Dolan:2014ska}.  The lower bounds are obtained to be  $g_p\gtrsim 3\times 10^{-7}, 5\times 10^{-7} v/\sqrt{m_t m_\chi}$, and $5\times 10^{-7} $ for the $CP$-violating interaction with quark universal couplings, quark Yukawa-like couplings, and quark first generation couplings, respectively, while  $g_p\gtrsim 2\times 10^{-7}$ \cite{Dolan:2014ska}, $3\times 10^{-7} v/\sqrt{m_t m_\chi}$ \cite{Dolan:2014ska}, and $g_p\gtrsim 3\times 10^{-7}$ for the $CP$-conserving interaction with quark universal couplings, quark Yukawa-like couplings, and quark first generation couplings, respectively. All the allowed parameter regions shown in Fig.~\ref{fig:FlavorBound} are above these lower bounds of $g_p$.

\vskip0.3cm
\noindent{\underline{\bf Big bang nucleosynthesis}:} 

Second, we discuss the BBN bound, where the pseudoscalar decays only to the SM particles because we consider $m_A<m_\chi$ in this work. If the pseudoscalars survived with a longer lifetime, for instance $\tau_A  \gtrsim 1$ second, the deficit of the neutrino distribution functions due to the insufficient thermalization weakens the weak interaction rates between proton and neutron and the freeze-out time thus becomes earlier, so that $n/p$ ratio as well as  $^4$He abundance becomes larger than in the standard BBN \cite{Kawasaki:2000en}. Following the result given in Ref. \cite{Kawasaki:2000en} that if the reheating temperature is larger than 0.7 MeV (corresponding to $t\lesssim 1$ second), the theoretical prediction of the $^4$He can remain within the 95\% CL limit of the observed abundance, we thus require that the lifetime of the pseudoscalar is less than 1 second.  We will be interested in the low-mass region $m_A\lesssim 3m_\pi$, where the constraint is stronger due to having a longer lifetime than the heavier one, and the only decay channel is $A\to \gamma\gamma$~\footnote{The pseudoscalar decays into pairs of leptons are irrelevant to the present work, since we consider that $A$ couples only directly to the quark sectors. For $m_A\lesssim 3\pi$, $A\to \pi\pi$ is forbidden by $CP$ symmetry, although it is kinematically allowed \cite{Dolan:2014ska}.}. The decay width is given by \cite{Gunion:1988mf,Djouadi:2005gj}
\begin{align}
\Gamma(A \rightarrow \gamma \gamma) & = \frac{\alpha^2 \, m_A^3}{64 \pi^3}\left|\sum_q  N_c \, Q_q^2 \, g_p^q \frac{f(\tau_q)}{m_q \tau_q}\right|^2 \; ,
\end{align}
where $\tau_q \equiv m_A^2 / (4 \, m_q^2)$, $N_c$ is the number of colors, $Q_q$ is the electromagnetic charge of the quark, and
\begin{equation}
f(\tau)  = \left\{ \begin{array}{lr} \text{arcsin}^2 \sqrt{\tau}
        \, , \ \ & \tau \leq 1 \\ -\frac{1}{4} \left[ \log \frac{1+\sqrt{1-\tau^{-1}}}{1-\sqrt{1-\tau^{-1}}} - i \pi\right]^2 \, , & \tau > 1 \end{array} \right. \; .
\end{equation}
We find that $g_p<10^{-8}$ is excluded in the low $m_A\lesssim 3m_\pi$ region for the cases of quark universal couplings and quark first-generation couplings, while for the case of quark Yukawa-like couplings the $g_p>10^{-4}$ region is allowed. The latter is shown in Fig. \ref{fig:FlavorBound}.

\vskip0.3cm
\noindent{\underline{\bf DM self-interaction}:}

Third, we consider the constraint due to DM self-interactions. The DM self-interactions can interpret  the small-scale structure of the Universe \cite{Spergel:1999mh}. To be consistent with astrophysical observations, the cosmological simulations have shown that  $\sigma/m_\chi\simeq  0.1 -10$ cm$^2/$g, where $\sigma$ is the DM self-interaction cross section \cite{Vogelsberger:2012ku,Rocha:2012jg,Zavala:2012us,Peter:2012jh}. It has been pointed out that the self-interactions of the DM mediated by a light dark force with the electromagnetic strength coupling can flatten the density profile around cores of dwarf galaxies \cite{Loeb:2010gj,Tulin:2013teo}. 

For a pseudoscalar that couples to the DM particle via the $CP$-violating coupling (${\cal L}_{\rm int}\supset   g_{s,\chi}    A \overline{\chi}  \chi$), the calculation for the DM self-interactions is completely the same as for the interactions arising from a scalar mediator with a $CP$-conserving coupling to the DM particle. 

 Under the conditions $g_{s,\chi}^2/(4\pi) \lesssim 10^{-2}$  and 7 GeV  $\lesssim m_\chi \lesssim $ 50 GeV, which are constrained by the DAMA data, and using the results given in Ref.~\cite{Tulin:2013teo}, we can place the bound for the pseudoscalar mass as 0.001~GeV $\lesssim m_A \lesssim$ 0.3~GeV.

For a pseudoscalar with a $CP$-conserving coupling to the DM particle (${\cal L}_{\rm int}\supset   g_{p,\chi}    A \overline{\chi} \gamma_5 \chi$), the calculation will be similar to the nucleon-nucleon interaction via one-pion exchange in the nuclear physics. However, we omit this part because the calculation is quite sophisticated  and a thorough treatment of it is beyond the scope of this paper.

\section{Summary}\label{sec:summary}

We have studied the fermionic DM particle interacting with the SM quarks via a light pseudoscalar mediator. Assuming that the $CP$ is not violated in the visible sector, we  separately consider the scenarios that the DM-pseudoscalar coupling is $CP$ conserving or $CP$ violating. 

Using the full form of interactions, we have shown that the replacement  ${g_{p(s),\chi} g_p}/({|\vec{q}|^2 +m_A^2}) \to 1/\Lambda^2$ is not suitable even when the mediator mass is of the same order of magnitude as the typical momentum transfer  at the direct-detection experiments, such that the allowed DAMA region is excluded or considerably modified by the correct relic density requirement (see Figs. \ref{fig:Relic-cpc} and \ref{fig:Relic-cpv}).

Considering the cases of quark universal couplings and Yukawa-like couplings,  only a small parameter region $m_A\lesssim$ 15 MeV can be accommodated for the latter, where the long-range interactions, instead of contact interactions, occur in the DM-iodine scatterings.
Our results seem to indicate that the fermionic DM-nucleus scattering mediated by a light pseudoscalar is dominated by the $CP$-violating interaction, i.e.,  $g_{s,\chi}  \gg 10^{-3} g_{p,\chi}$  (see Figs. \ref{fig:Relic-cpc} and \ref{fig:Relic-cpv}).

We find that the interference between the term containing $F_{\Sigma''}^{(p,n)}$ and that containing $F_{\Sigma''}^{(p,p)}$ and  $F_{\Sigma''}^{(n,n)}$ is destructive for $c_p/c_n<0$. Especially for $c_p/c_n\approx -60\sim -40$, the exclusion limits set by  SuperCDMS, XENON100, and LUX are highly suppressed, and the DAMA signal can thus be easily reconciled with these null measurements (see Figs. \ref{fig:DirectDetectCPC} and \ref{fig:DirectDetectCPV}).

For this fermionic DM simplified model, the allowed region set by the DAMA signal and the correct relic density can successfully satisfy  the  conditions requiring by the thermal equilibrium, big bang nucleosynthesis, and DM self-interactions. Most DAMA regions may be excluded by flavor constraints for quark universal couplings and Yukawa-like couplings (see Fig.~\ref{fig:FlavorBound}).
However, because  the simplified model  is a model-independent  bottom-up approach, a phenomenological extension  of this model can change the values of FCNCs; in other words, the present flavor constraints may be overestimated. Nevertheless, the results of future measurements on flavor physics will still provide important constraints on the related models. In addition, more precise measurements performed by COUPP, PICASSO, SIMPLE, and KIMS, which contain target nuclei with unpaired protons as the DAMA experiment, will offer more information to  test this model. Thus, the tension between the DAMA results and these measurements could be clarified.

\acknowledgments \vspace*{-1ex}
 This work was supported in part by the Ministry of Science and Technology of the Republic of China under Grant No. 102-2112-M-033-007-MY3.

\end{document}